\documentclass[structabstract]{aa}  
\usepackage{txfonts,graphicx,natbib}
\begin{document}
\title{Large-scale mapping of the massive star-forming region RCW38 in the [CII] and PAH emission}
\titlerunning{[C{\small II}] and PAH mapping of RCW38}
\authorrunning{H. Kaneda et al.}
\author{H. Kaneda\inst{1}, T. Nakagawa\inst{2}, S. K. Ghosh\inst{3}, D. K. Ojha\inst{3}, D. Ishihara\inst{1}, T. Kondo\inst{1}, J. P. Ninan\inst{3}, M. Tanabe\inst{1}, Y. Fukui\inst{1}, Y. Hattori\inst{1}, T. Onaka\inst{4}, K. Torii\inst{1}, \and M. Yamagishi\inst{1}}
\institute{
Graduate School of Science, Nagoya University, Chikusa-ku, Nagoya, 464-8602, Japan \\
\email{kaneda@u.phys.nagoya-u.ac.jp}\\
\and
Institute of Space and Astronautical Science, Japan Aerospace Exploration Agency, Sagamihara, Kanagawa 252-5210, Japan\\
\and
Tata Institute of Fundamental Research, Homi Bhabha Road, Mumbai 400005, India \\
\and
Department of Astronomy, Graduate School of Science, University of Tokyo, \\
Bunkyo-ku, Tokyo 113-0033, Japan
}

\date{Received; accepted}

\abstract
  % context heading (optional)
  % {} leave it empty if necessary  
{}
  % aims heading (mandatory)
{We investigate the large-scale structure of the interstellar medium (ISM) around the massive star cluster RCW~38 in the [C{\small II}] 158 $\mu$m line and polycyclic aromatic hydrocarbon (PAH) emission.}
  % methods heading (mandatory)
{We carried out [C{\small II}] line mapping of an area of $\sim30'\times15'$ for RCW~38 by a Fabry-Perot spectrometer on a 100 cm balloon-borne telescope with an angular resolution of $\sim1'.5$. We compared the [C{\small II}] intensity map with the PAH and dust emission maps obtained by the AKARI satellite.}
  % results heading (mandatory)
{The [C{\small II}] emission shows a highly nonuniform distribution around the cluster, exhibiting the structure widely extended to the north and the east from the center. The [C{\small II}] intensity rapidly drops toward the southwest direction, where a CO cloud appears to dominate. We decompose the 3--160 $\mu$m spectral energy distributions of the surrounding ISM structure into PAH as well as warm and cool dust components with the help of 2.5--5 $\mu$m spectra.  } 
{We find that the [C{\small II}] emission spatially corresponds to the PAH emission better than to the dust emission, confirming the relative importance of PAHs for photo-electric heating of gas in photo-dissociation regions. A naive interpretation based on our observational results indicates that molecular clouds associated with RCW~38 are located both on the side of and behind the cluster.}
\keywords{ISM: dust, extinction -- infrared: ISM: lines and bands -- ISM: structure -- ISM: clouds}

\maketitle

\section{Introduction}
The structure of the interstellar medium (ISM) around massive star clusters provides crucial information on their formation environment in the past as well as their interplay with ambient clouds at present. According to its definition, a photo-dissociation region (PDR) is pervasively formed around an HII region created by a star cluster, where hydrogen gas is neutral (either atomic or molecular), molecular oxygen is photo-dissociated, and far-ultraviolet (far-UV; 6--13.6 eV) fluxes are important in gas heating and chemistry (\citealt{Tie85}). A classical concept defines a PDR simply as a transition zone between an HII region and a molecular cloud. In reality, PDRs reveal internal fine structures such as filaments or clumps (e.g., \citealt{Goi11}), and the ISM in various gas phases, which include HII regions and molecular clouds, can be entangled heavily along a line of sight. 

In PDRs, dust and polycyclic aromatic hydrocarbons (PAHs) absorb stellar far-UV photons and re-radiate photons in the infrared (IR), where PAHs are responsible for $\sim10$ \% of the re-emitted IR fluxes. About 0.1--1 \% of the far-UV energy is converted to photoelectrons that are ejected from dust and PAHs to heat the gas (e.g., \citealt{Hol99}). Because PAHs are the smallest forms of dust grains with a high number density, and thus provide a large surface area per unit volume, they are believed  to be the most important gas heating agents (\citealt{Bak94}; \citealt{Wei01}; \citealt{Hab01}). Gas cools in fine-structure forbidden lines, among which [C{\small II}] is the most important gas coolant in low-density PDRs (e.g., \citealt{Hol99}). PAHs in an electrically neutral state emit much less at 6--9 $\mu$m than those in an ionized state (\citealt{Szc93}; \citealt{Hud95}; \citealt{All99}; \citealt{Dra07}); in regions where neutral PAHs are dominant, PAH-ionizing photons are few and accordingly photo-electrons are not available to heat the gas. Thermal emission of large grains comes not only from PDRs but also from HII regions and molecular clouds. There is, however, a significant difference in equilibrium temperature: $30\sim50$ K for HII regions and the surfaces of PDRs, and $\lesssim20$ K for the others (e.g., \citealt{Sod87, Hol91, Sod94}). Thus a combination of the [C{\small II}] 158 $\mu$m line, ionized and neutral PAH features, and warm and cool dust emission is of great use in disentangling the ISM in various gas phases. In particular, a strong correlation is expected in a PDR between the far-IR [C{\small II}] line emission and the PAH emission in the near- to mid-IR. For example, \citet{Hel01} reported a strong correlation between the [C{\small II}] line and the ISOCAM 5--10 $\mu$m flux for star-forming galaxies. Until very recently, however, spatial resolutions in the far-IR were extremely poor compared with those in the near- and mid-IR, which hampered a detailed comparison of spatial distributions of the [C{\small II}] and PAH emission. With the advent of Herschel and SOFIA, this situation is now improving significantly (e.g., \citealt{Job10}, \citealt{Oka12, Oka13}), but spatial coverage large enough to map the whole structures of nearby massive star-forming regions is still neither easy nor effective with large space telescopes.

In this paper, we present the results of large-scale mapping of the massive star-forming region RCW~38 in the [C{\small II}] line and PAH emission. The RCW~38 region contains a large star cluster with thousands of stars, including an O5 star IRS 2 (\citealt{Fro74}). \citet{Der09} identified more than 300 young stars in the $J$, $H$, and $K_s$ bands for the $\sim$0.5 pc$^2$ area centered on the IRS2. In fact, within 2 kpc from Sun, the Orion Nebula Cluster and RCW~38 are the two largest star clusters (\citealt{Wol06}). The star cluster itself has been studied in many papers (e.g., \citealt{Smi99}, \citealt{Win11}). This paper focuses on the ambient ISM widely distributed around the cluster. Large-scale $^{13}$CO surveys with the NANTEN telescope showed that two molecular clouds are associated with RCW38, one positionally coincident with IRS2 and the other widely extending to the west (\citealt{Yam99}).   
Because RCW~38 is thought to be less evolved and thus more embedded than the Orion, it is ideal to study the condition of the ISM associated with such a massive cluster.  The data of the [C{\small II}] line emission were taken with a Fabry-Perot spectrometer aboard a 100 cm balloon-borne far-IR telescope (FPS100; \citealt{Gho88}; \citealt{Nak98}), while the data of the PAH and dust emission were taken by all-sky surveys with the Infrared Camera (IRC; \citealt{Ona07}) and the Far-Infrared Surveyor (FIS; \citealt{Kaw07}) aboard the AKARI satellite (\citealt{Mur07}). We adopt below a distance of 1.7 kpc to RCW~38 (Beck et al. 1991). As shown below, a combination of the FPS100 observation with the AKARI survey enables us to make a comparison of large-scale ISM structures in the [C{\small II}] and PAH emission around RCW~38. 

\section{Observation and Data reduction}
The [C{\small II}] observation of RCW~38 was carried out on 05 Feb. 2009 by the balloon-borne FPS100, which was flown from the Hyderabad Balloon Facility of the Tata Institute of Fundamental Research (TIFR) in India. The payload was released at 23:00:46 (Indian Standard Time; IST = UT$+5.5$ h) and RCW~38 was observed from 26:34 to 27:25 (IST) by spatial raster scans covering a rectangular area of $\sim30'\times15'$ around the center of RCW~38: R.A. $=$ 08 59 05.5 and Dec. $=$ $-$47 30 39 (J2000). Prior to the observation, the telescope was pointed toward Saturn for absolute flux calibration. The description of the FPS100 and its observation modes were given in Mookerjea et al. (2003). A spatially unchopped, fast spectral scan mode was used in the present observation.  

Figure 1a shows an example of the [C{\small II}] line profiles taken during the observation of RCW~38 with the FPS100, where atmospheric background and astronomical continuum components are subtracted. An atmospheric background spectrum is obtained by averaging the spectral scan data at both edges of the spatial scan legs every few scans; the spatial scan length is $\sim30'$ and the first and the last $5'$ regions are used to estimate the background, where there is no apparent contribution from RCW~38. Then we fitted the data of the corresponding spatial scans by a combination of the atmospheric component, a linear function representative of an astronomical continuum component, and a Lorentzian profile of the [C{\small II}] line emission. Here the normalizations of the three components were allowed to vary, while the center and the width of the Lorentzian were fixed to the values pre-determined by using the data of several spectral scans with high signal-to-noise ratios. 
We made a fine tuning of the positional offset in the map ($-6\fs 4$ in R.A., $-19\farcs 8$ in Dec.), matching the central peak in the continuum emission with that in the IRAS 100 $\mu$m band. 
The resultant [C{\small II}] intensity map is shown in Fig.1b, which has a $30''$ grid spacing. The contours show the intensity levels of $2.1\times 10^{-4}$ to $3.4\times 10^{-3}$ ergs s$^{-1}$ cm$^{-2}$ sr$^{-1}$, while the 1-sigma error of the intensities is typically $\sim 4\times 10^{-5}$ ergs s$^{-1}$ cm$^{-2}$ sr$^{-1}$ estimated from errors of the spectral fitting. Hence the map shows a significant detection of widely extended [C{\small II}] emission associated with RCW~38. We find that the [C{\small II}] emission of RCW~38 exhibits a rather complicated structure. In an inner ($<3'$) region, the emission is elongated along the northeast (NE) --southwest (SW) direction. In an outer region, the emission is extended widely ($\sim10'$) to the north and the east from the center, while it falls off sharply toward the SW direction. 

For comparison with the [C{\small II}] map, in Fig.2 we show the maps of RCW~38 obtained by the AKARI all-sky-survey in the {\it S9W} (the reference wavelength and band width of 9 $\mu$m and 6.7--11.6 $\mu$m; \citealt{Ona07}) and {\it L18W} (18 $\mu$m and 13.9--25.6 $\mu$m) wide bands, and the {\it N160} (160 $\mu$m and 145--180 $\mu$m; \citealt{Kaw07}) narrow band, which are expected to be dominated by the emission of ionized PAHs, warm dust, and cool dust, respectively. As for the areas hatched in grey, the {\it L18W} map shows a signal saturation near the center, while a small portion of the {\it N160} map is masked because this band is known to exhibit a ghost of the central peak at the hatched position. The {\it WIDE-S} (90 $\mu$m and 75--110 $\mu$m) and {\it WIDE-L} (140 $\mu$m and 110--170 $\mu$m) band data cannot be used for the other bands in the all-sky survey data, because they are severely saturated in a large fraction of the central region due to their wide band widths. In contrast, the {\it N60} (65 $\mu$m and 55--75 $\mu$m) narrow band data are used in the discussion later, although a small portion of the central region is saturated. In addition, AKARI conducted pointed imaging observations of a central $10'\times10'$ area of RCW~38 in the {\it N3} (3.2 $\mu$m and 2.7--3.8 $\mu$m; \citealt{Ona07}), {\it N4} (4.1 $\mu$m and 3.6--5.3 $\mu$m), {\it S7} (7.0 $\mu$m and 5.9--8.4 $\mu$m), and {\it S11} (11.0 $\mu$m and 8.5--13.1 $\mu$m) narrow bands, along with spectroscopic observations at 2.5--5.0 $\mu$m (\citealt{Ohy07}). Because the pointed-observation maps have a better imaging quality but a smaller spatial coverage than the all-sky-survey maps, we use the former and the latter maps for discussions on inner small-scale and outer large-scale structures, respectively. The log of the AKARI pointed observations of RCW~38 (and RCW~49, which will be used as a reference sample) is summarized in Table 1. 

\begin{figure*}
   \centering
   \includegraphics[width=6cm]{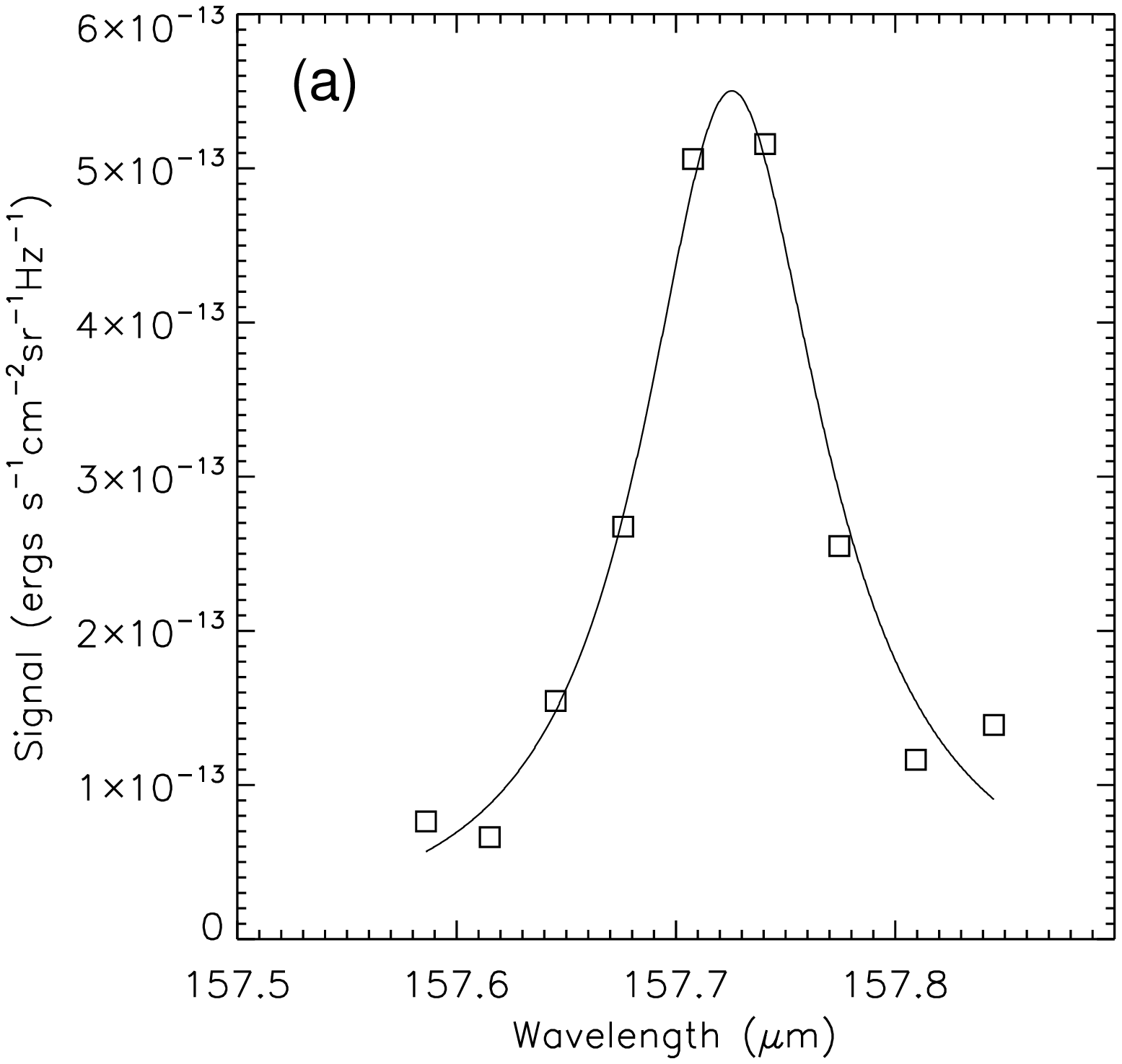}
   \includegraphics[width=7.5cm]{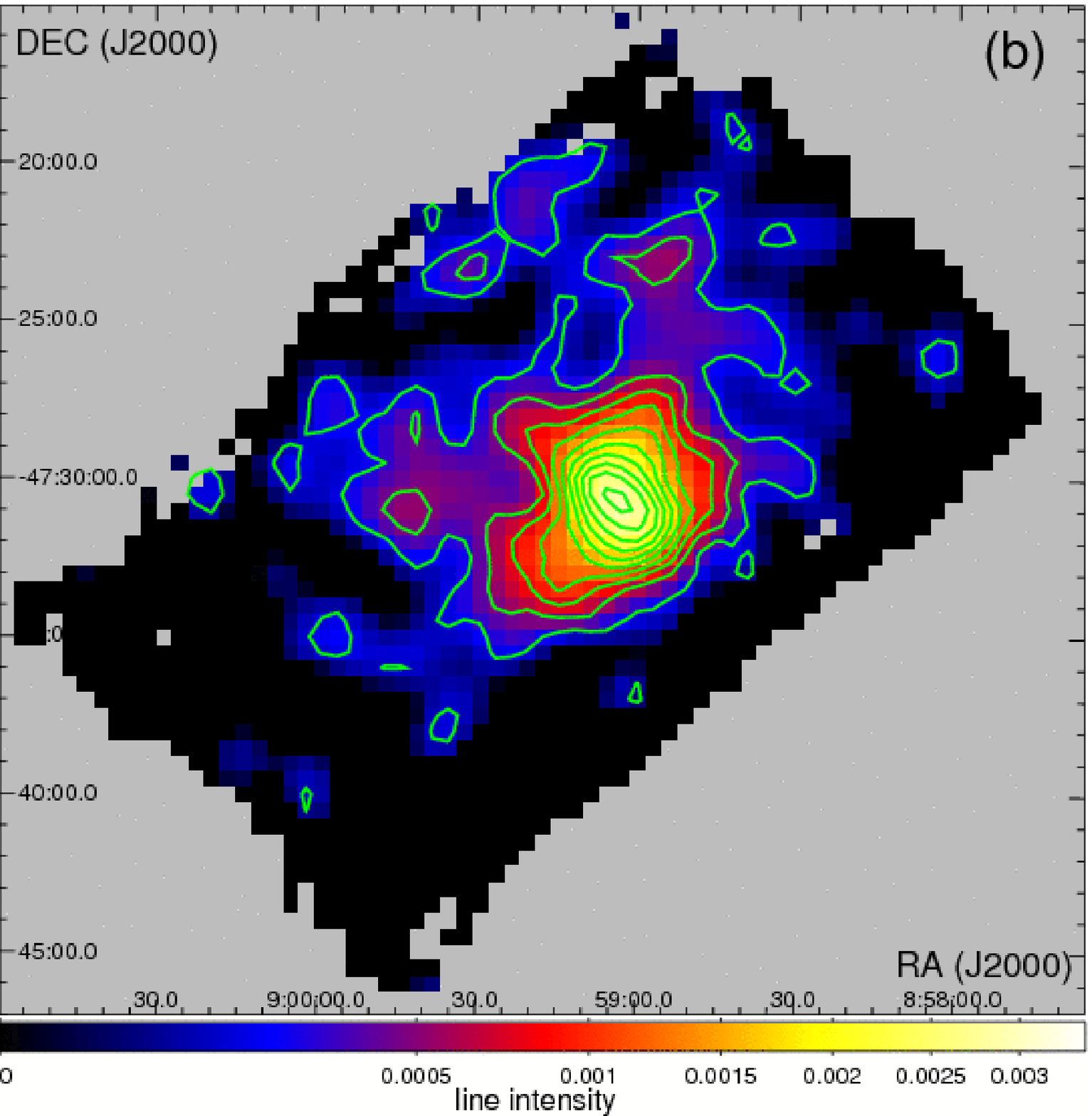}
   \caption{(a) [C{\small II}] line profile in a single spectral scan obtained during the observation of RCW~38 with the balloon-borne FPS100. (b) [C{\small II}] line map of an area of about $30'\times 15'$ around RCW38. The color levels are given in units of ergs s$^{-1}$ cm$^{-2}$ sr$^{-1}$. The contours are drawn with linearly spaced 10 levels from $2.1\times 10^{-4}$ to $3.4\times 10^{-3}$ ergs s$^{-1}$ cm$^{-2}$ sr$^{-1}$.}
   \label{}
\end{figure*}

\begin{figure*}
   \centering
   \includegraphics[width=6.5cm]{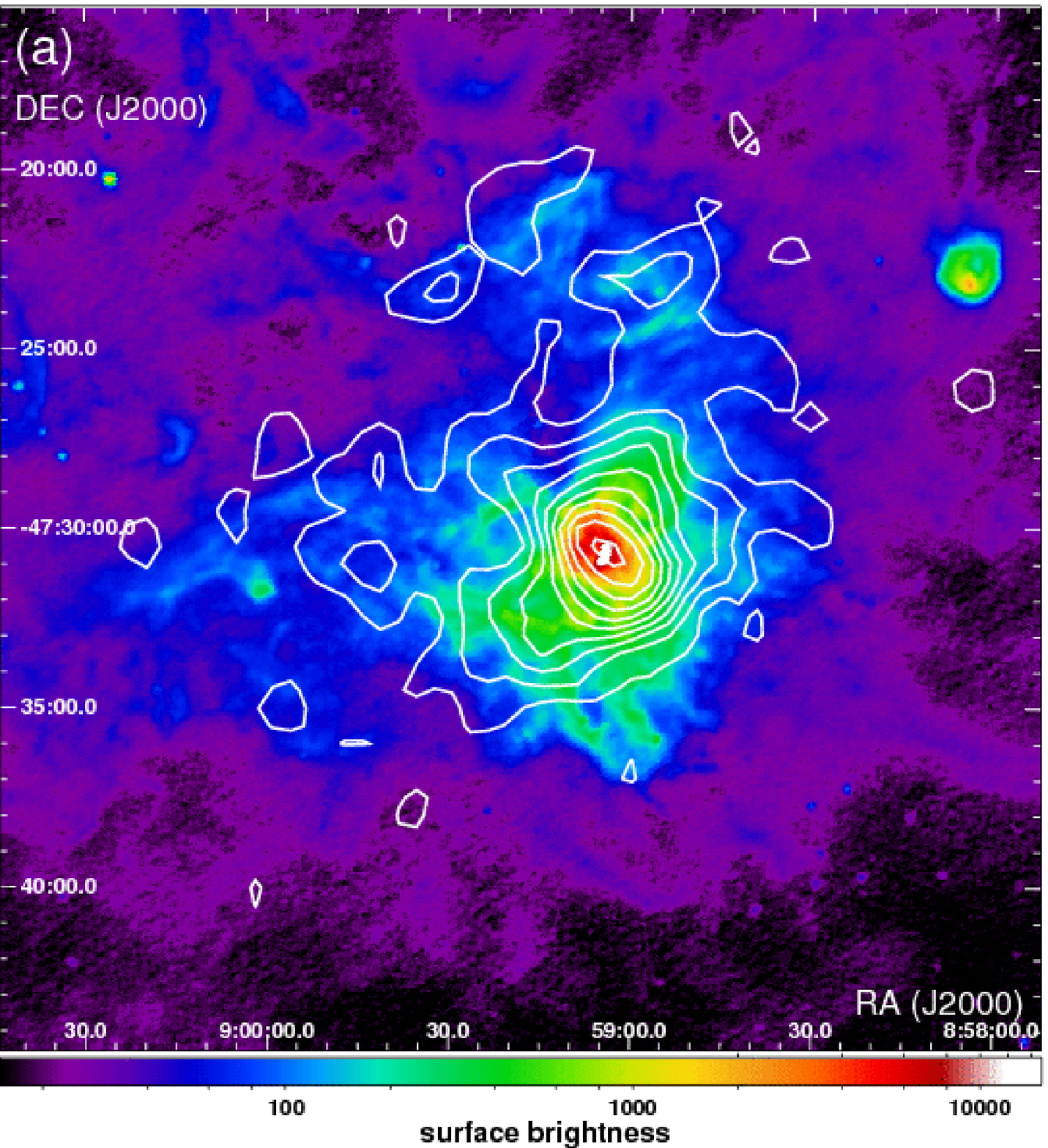}
   \includegraphics[width=6.5cm]{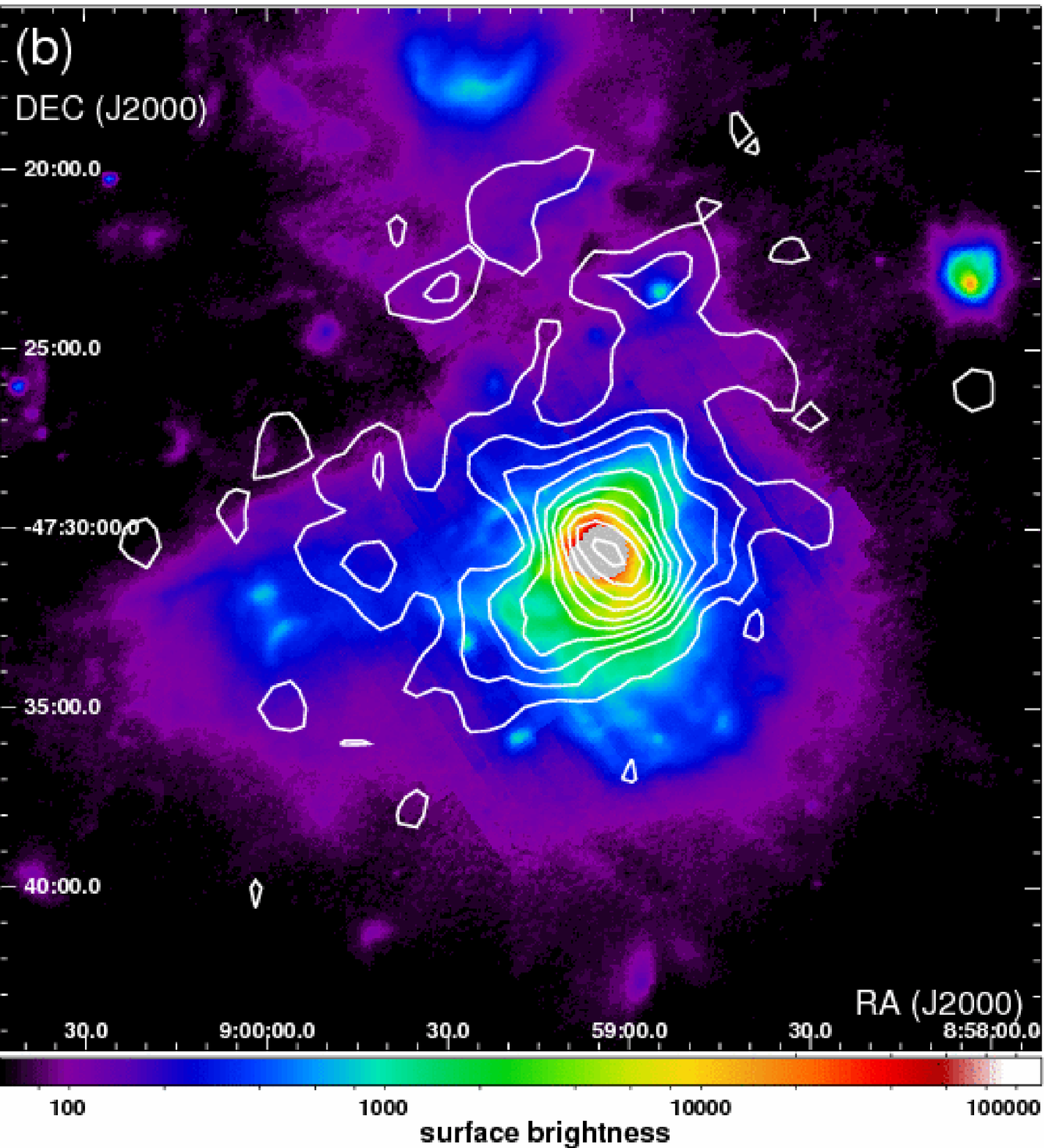}
   \includegraphics[width=6.5cm]{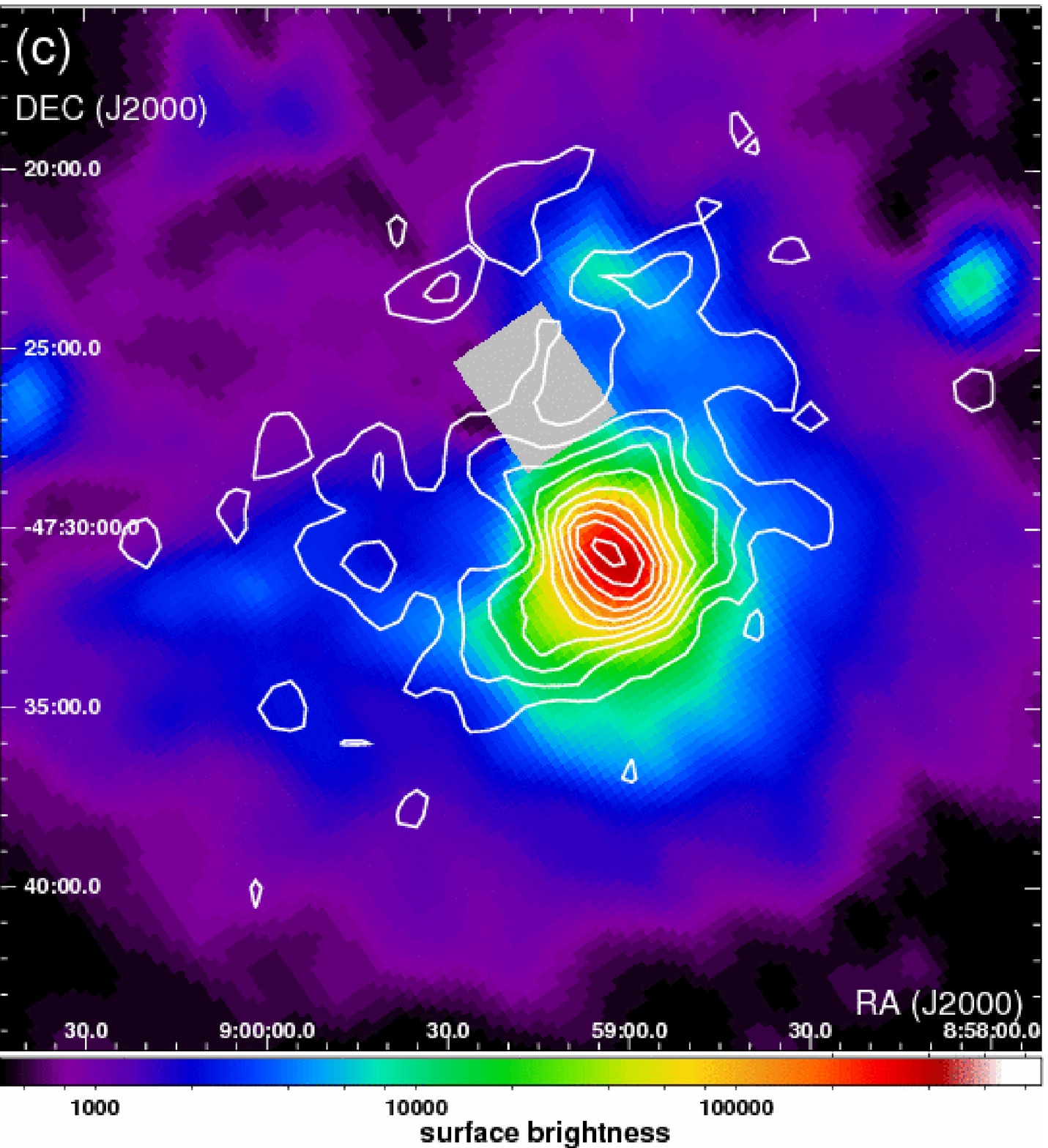}
   \includegraphics[width=6.5cm]{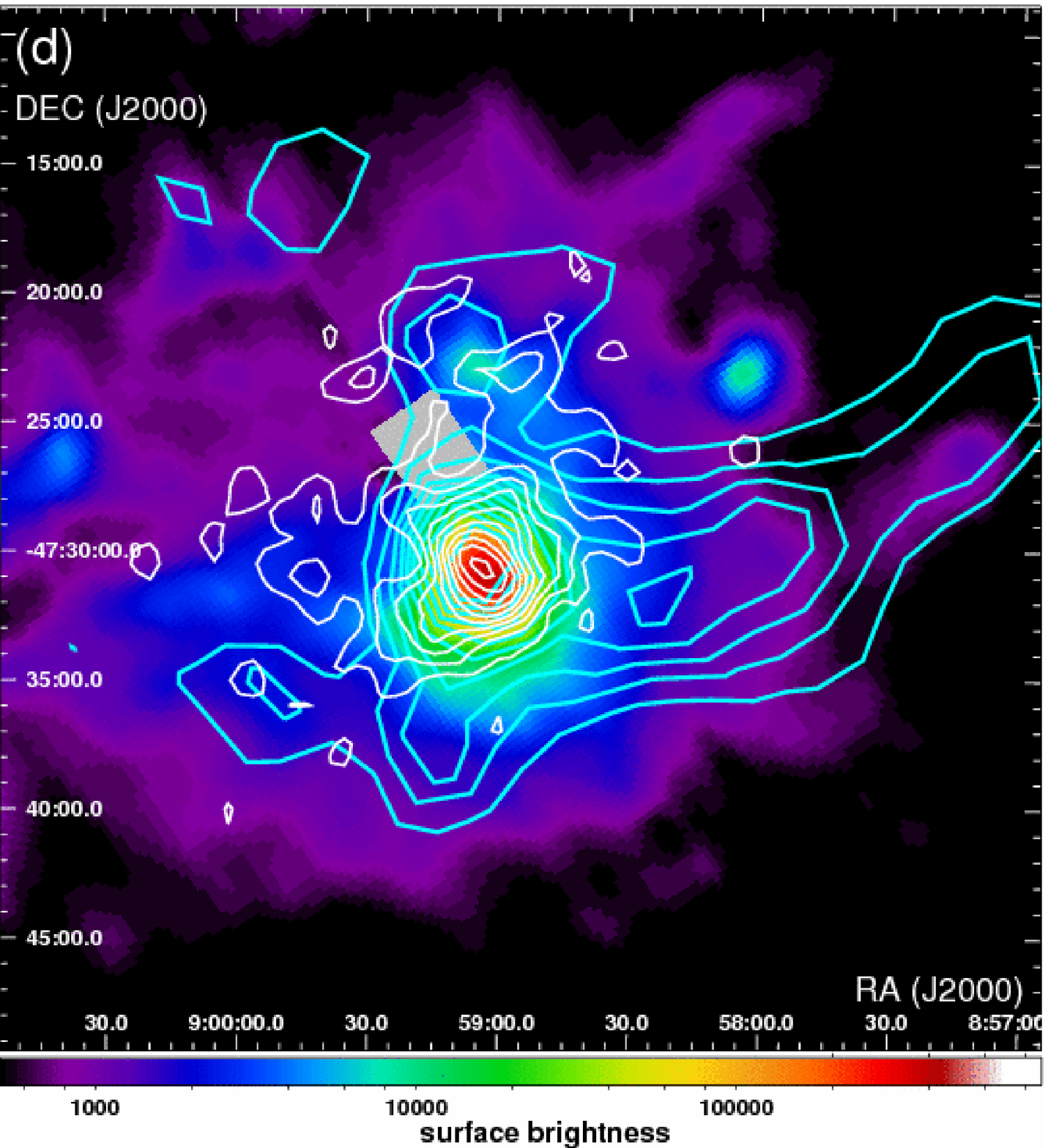}
   \caption{(a) AKARI {\it S9W} (9 $\mu$m), (b) {\it L18W} (18 $\mu$m), (c) {\it N160} (160 $\mu$m) band maps of RCW38, shown together with the [C{\small II}] contour map in Fig.1. The color levels are given in units of MJy sr$^{-1}$. The AKARI maps are derived from the all-sky survey data. (d) Same as panel (c), but enlarged and compared with the NANTEN $^{12}$CO J=1-0 contours (velocity range: $-3$--$+12$ km s$^{-1}$), where the contour levels are linearly drawn from 30 to 100 K km s$^{-1}$ in 10 steps.  }
   \label{}
\end{figure*}

\begin{table*}
\caption{Observation log}
\label{log}
\centering
\renewcommand{\footnoterule}{}
\begin{tabular}{cccccc}
\hline\hline
Mode & R.A. (J2000) & Dec. (J2000) & Observation ID & Date \\
\hline
IRC imaging (RCW~38) & 08 59 00.0 & $-$47 30 34.9 & 1600044 & 03 Dec 2006 \\
IRC spectroscopy (RCW~38)& 08 59 04.2 & $-$47 30 15.5 & 1420754 & 01 Jun 2008 \\
IRC spectroscopy (RCW~49)& 10 24 12.6 & $-$57 46 28.6 & 1420552 & 06 Jul 2008 \\
\hline
\end{tabular}
\end{table*}

\section{Results}
For larger scale structures, the observation reveals that the [C{\small II}] emission extends far toward the east and the north from the center of RCW~38, while it falls off sharply toward the SW direction. As can be seen in Fig.2a, the distribution of the PAH emission follows that of the [C{\small II}] emission very well, suggesting the importance of PAHs for photo-electric heating of gas in PDRs. Figure 2b shows the distribution of mid-IR warm dust emission, which exhibits a significant difference in the large-scale structure from those of the [C{\small II}] and PAH emission. In the warm dust emission, the north extension can hardly be recognized, while the east extension is clear, suggesting the difference in the properties of the ISM extending in these directions. This also indicates that the dust grains are less important in the photo-electric heating of gas, at least in the north extension. As can be seen in Fig.2c, the distribution of the far-IR dust emission also does not follow well the [C{\small II}] emission in the SW region, where the far-IR dust emission is more extended.
Figure 2d shows the distribution of the $^{12}$CO J=1--0 emission observed with NANTEN, where the velocity range is $-3$ to $+12$ km s$^{-1}$. For the central $2'\times2'$ area, Gyulbudaghian \& May (2008) showed that the two clouds of $^{12}$CO J=1--0 emission with the velocity ranges of $-3$--$+2$ km s$^{-1}$ and $+3$--$+8$ km s$^{-1}$ are associated with RCW~38 from the Swedish-ESO Submillimetre Telescope (SEST) observation. Therefore the CO cloud shown in Fig.2d, as a whole, is most likely associated with RCW~38. On a large spatial scale, as already reported in \citet{Yam99} and depicted in Fig.2d, molecular clouds are positionally coincident with IRS2 and also lie predominantly in the SW region of RCW~38, which extends widely to the west. The latter component is likely to cause the rapid decrease of the [C{\small II}] emission in this direction. This demonstrates the fact that the star cluster adjoins very closely to the side of the molecular cloud, which is dense enough to prevent intense far-UV radiation of the cluster from penetrating deep inside the clouds.

Figure 3 shows correlation plots between 9 $\mu$m and [C{\small II}], 160 $\mu$m and [C{\small II}], and 160 $\mu$m and 9 $\mu$m surface brightness, where all the images are regridded to a common spatial scale of $90''\times90''$. The correlation coefficients are 0.85, 0.79, and 0.93 ($N=106$), respectively, and are calculated for their logarithmic intensities in the area where the [C{\small II}] intensity is higher than $4 \times 10^{-5}$ ergs s$^{-1}$ cm$^{-2}$ sr$^{-1}$. Hence all of them are significantly correlated with one another; the 9 $\mu$m and 160 $\mu$m intensities have the strongest correlation, which indicates that PAHs and large grains are mixed well in the ISM, as pointed out by many authors (e.g., \citealt{Ona96}; \citealt{Kan12}). It is apparent from Fig.2c that the degradation of the correlation between the [C{\small II}] and the 160 $\mu$m intensity is caused by the difference in spatial distribution in the SW region, where the 160 $\mu$m intensity is more extended. For high surface brightness areas, the 9 $\mu$m and the 160 $\mu$m brightness exceed their linear relation with the [C{\small II}] brightness. They indicate that the photo-electric heating efficiencies of both PAHs and large grains decrease due to their charge-up effects or that the [C{\small II}] line intensity is saturated at gas densities higher than its critical density (e.g., \citealt{Hol99}). An additional reason can be that PAHs are destroyed under intense UV irradiation (e.g., \citealt{Ber12}) and hence heating efficiency is reduced. 

\begin{figure*}
   \centering
   \includegraphics[width=6.5cm]{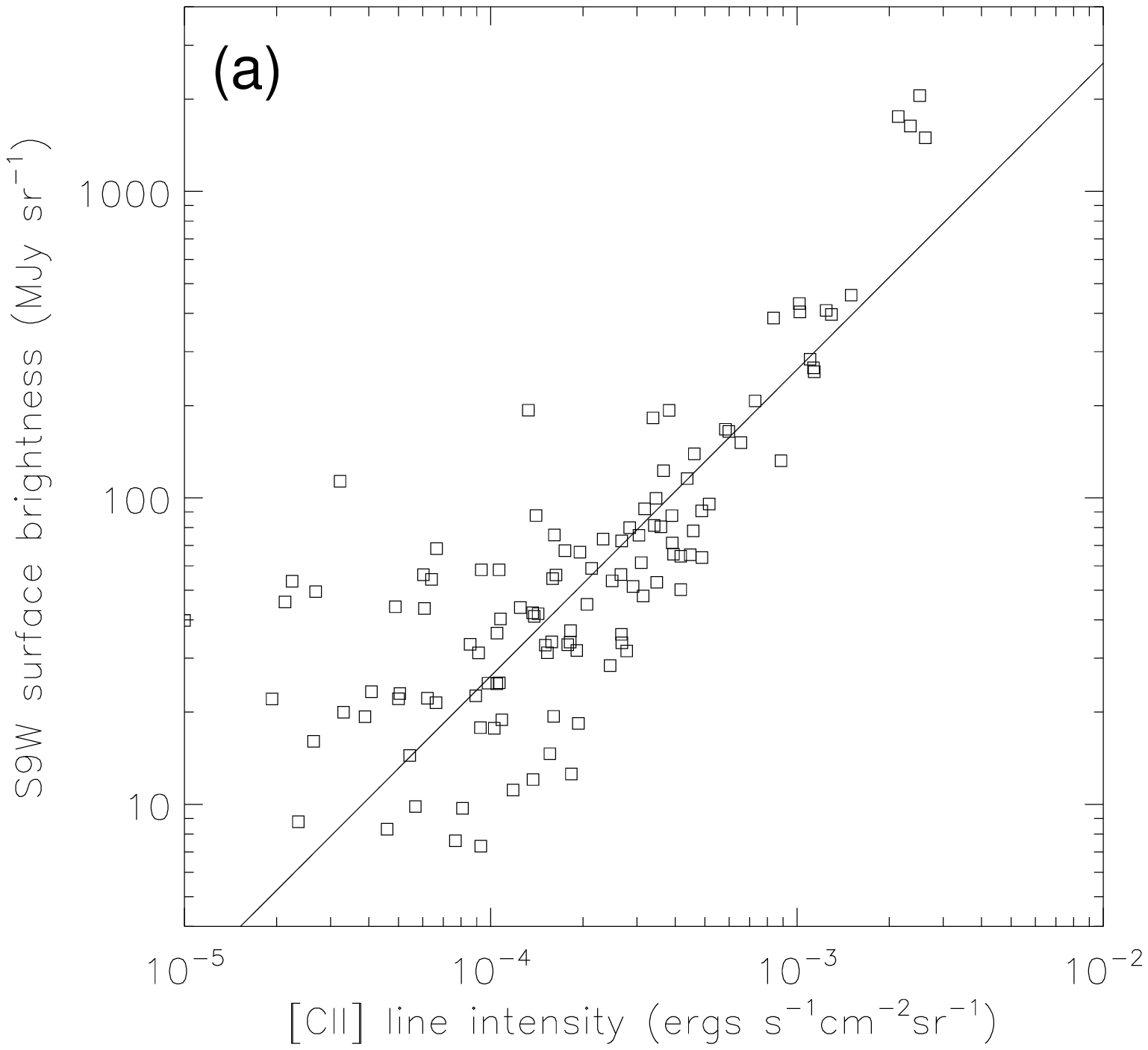}
   \includegraphics[width=6.5cm]{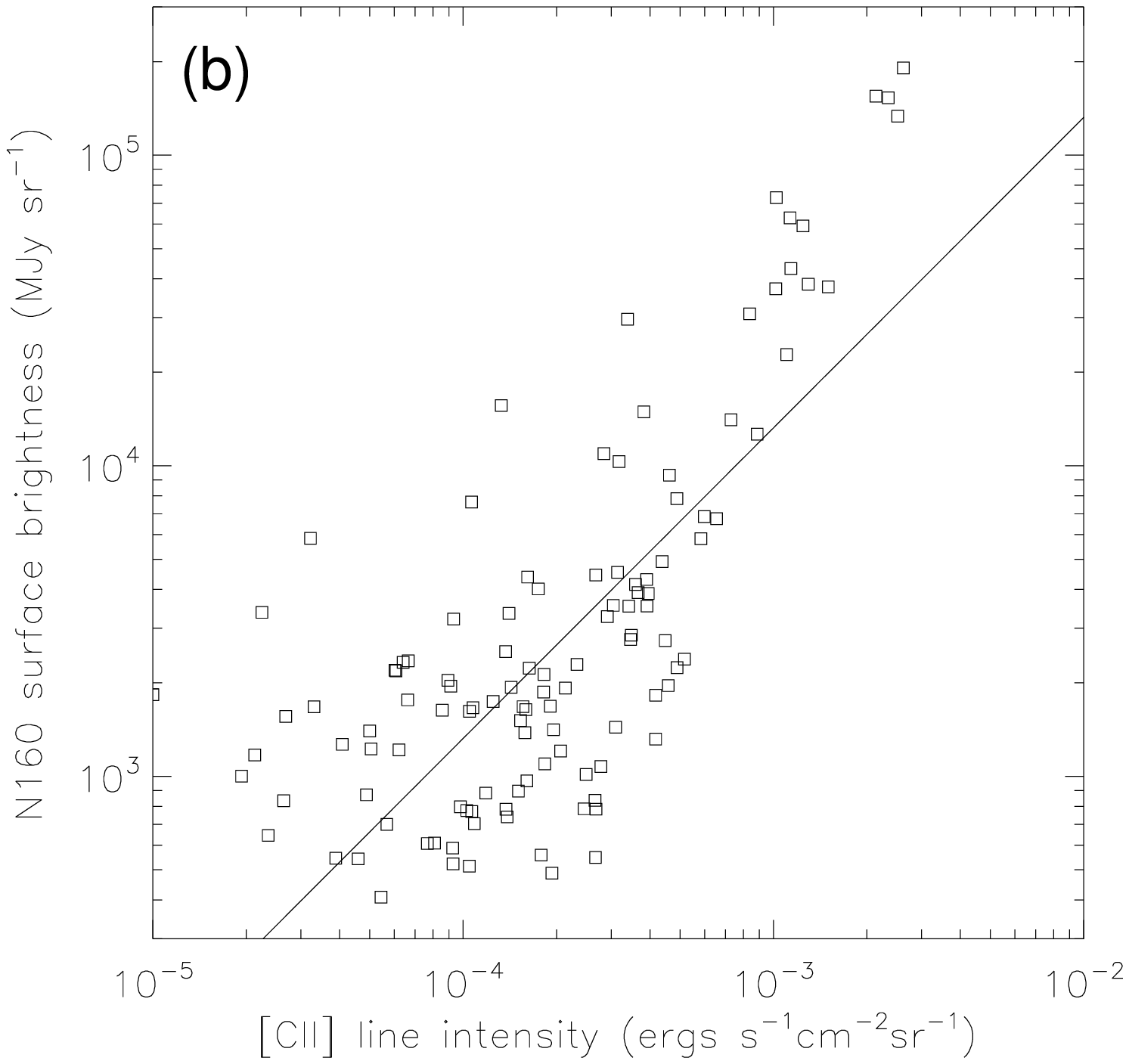}
   \includegraphics[width=6.5cm]{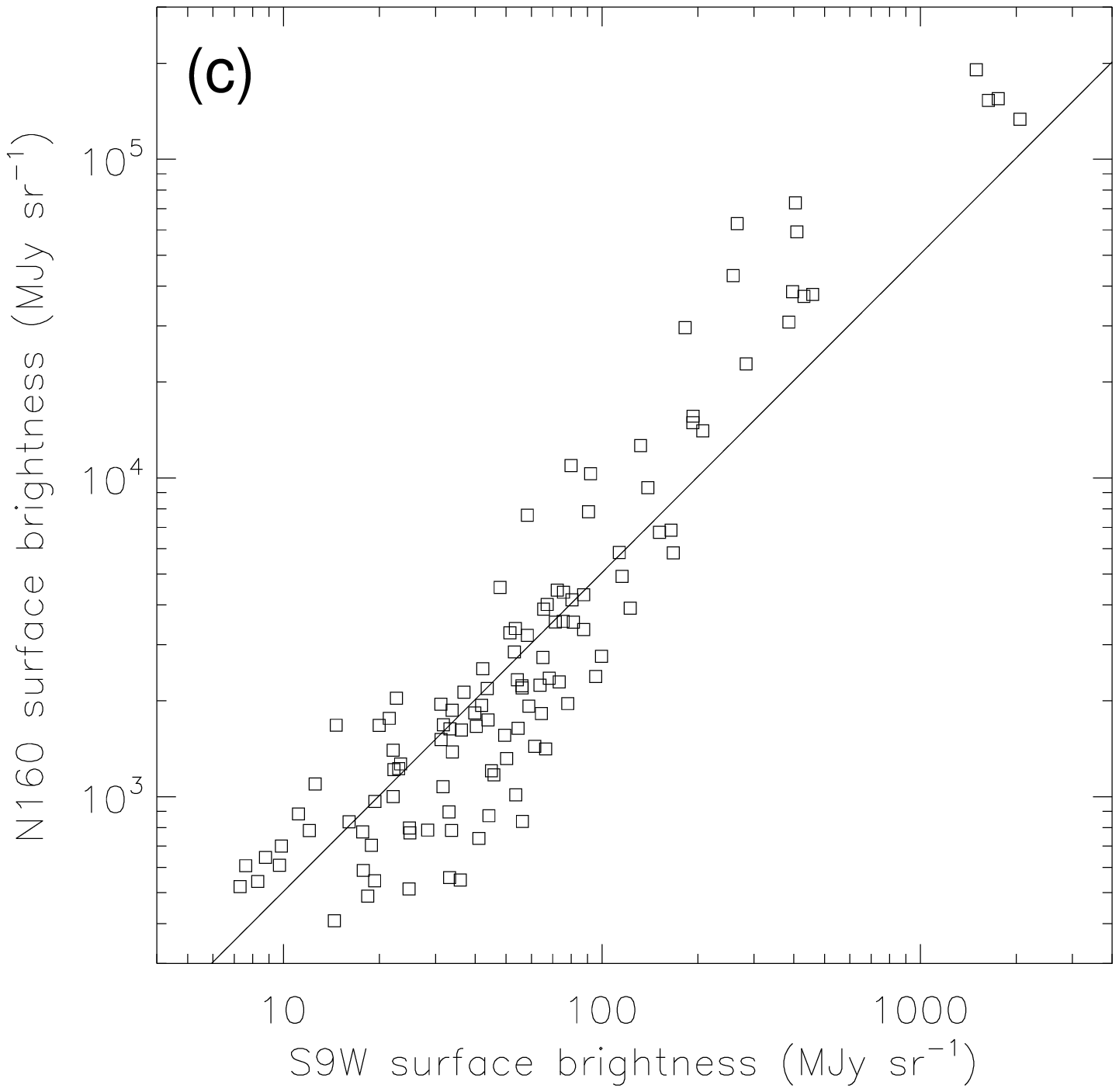}
   \caption{Correlation plots of (a) {\it S9W} (9 $\mu$m) versus [C{\small II}], (b) {\it N160} (160 $\mu$m) versus [C{\small II}], and (c) {\it N160} versus {\it S9W} brightness. All the data are plotted for ranges of three orders of magnitude. The data sample is the same among the plots.}
   \label{}
\end{figure*}

\begin{figure*}
   \centering
   \includegraphics[width=6.5cm]{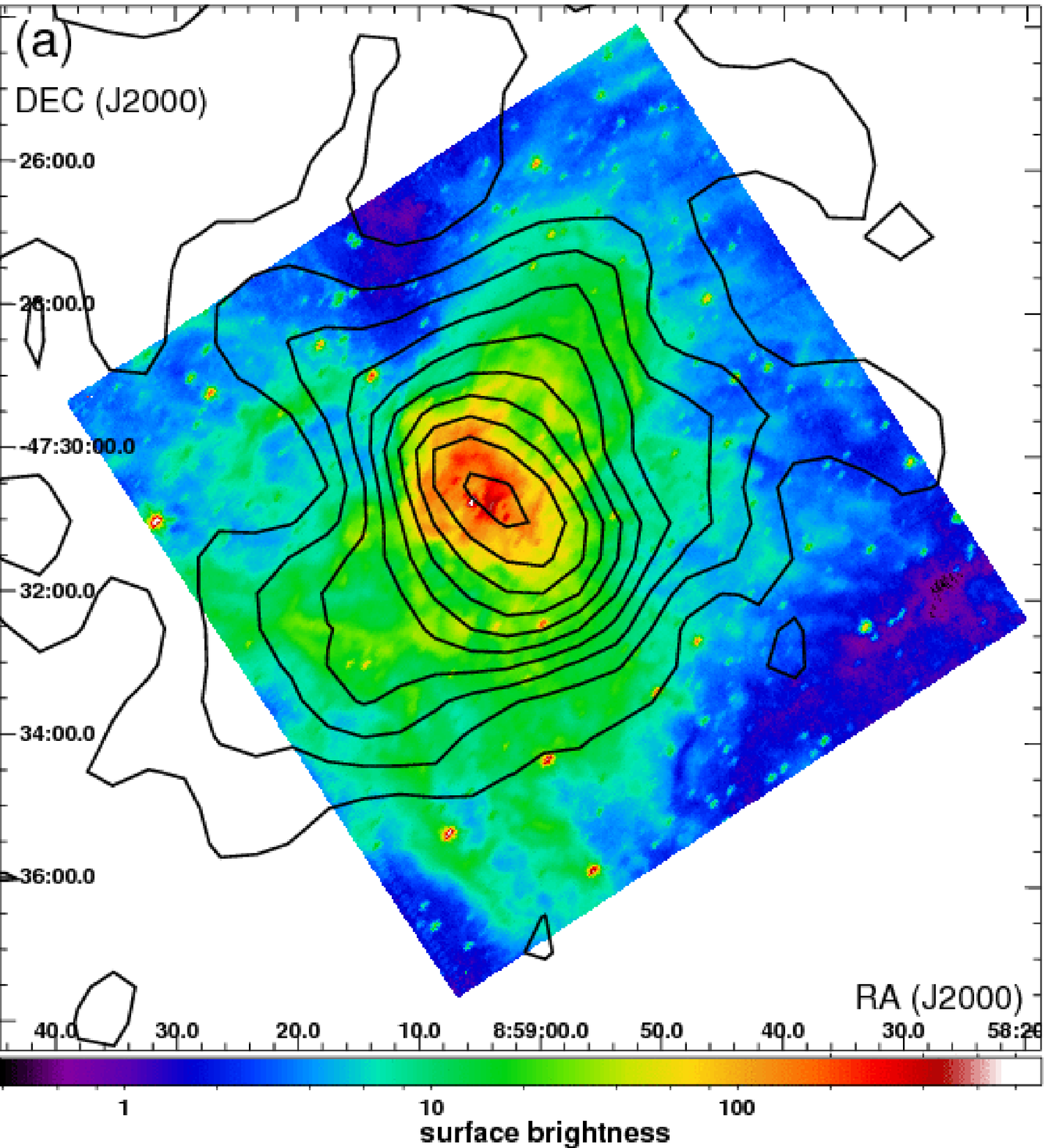}
   \includegraphics[width=6.5cm]{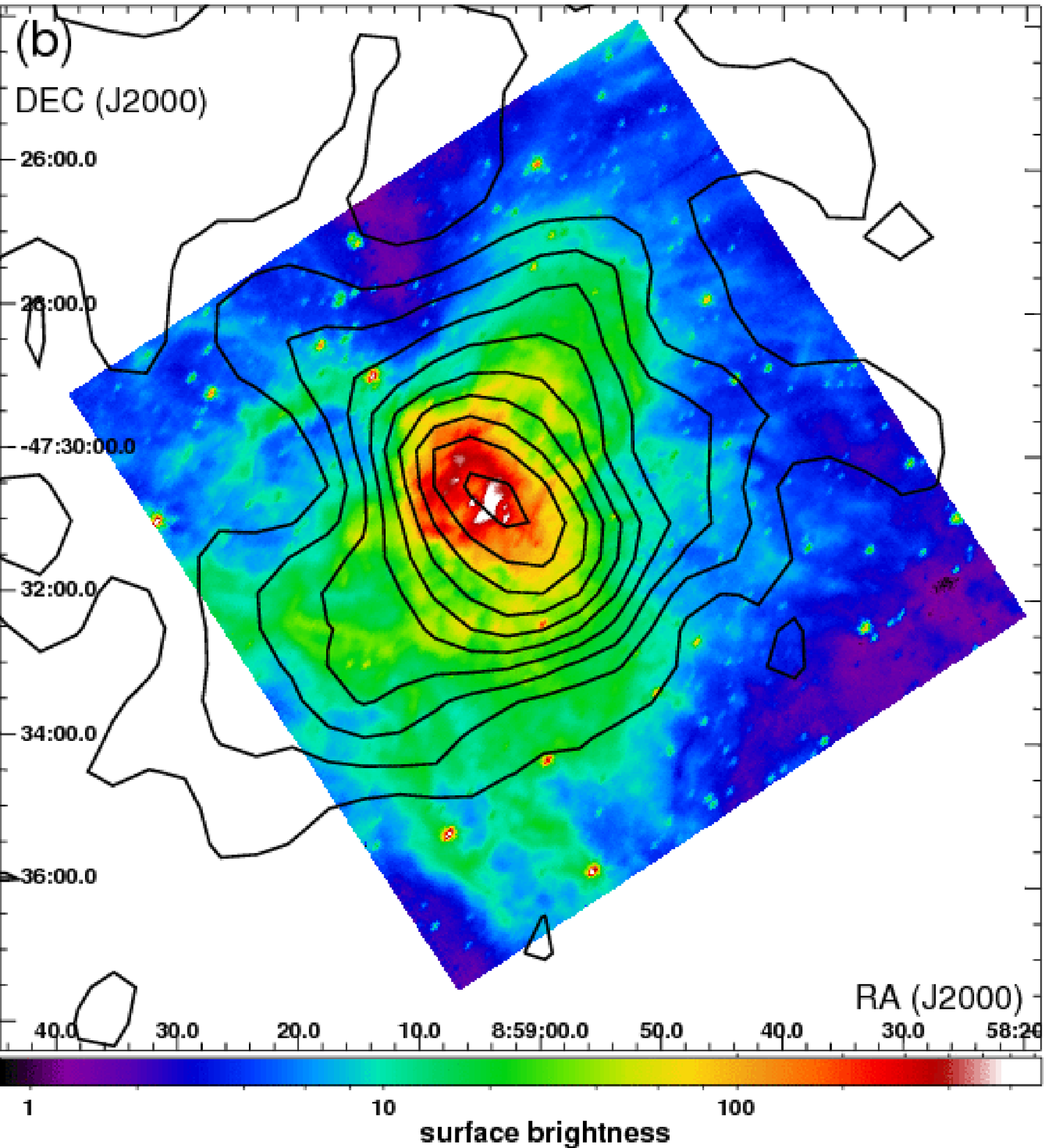}
   \includegraphics[width=6.5cm]{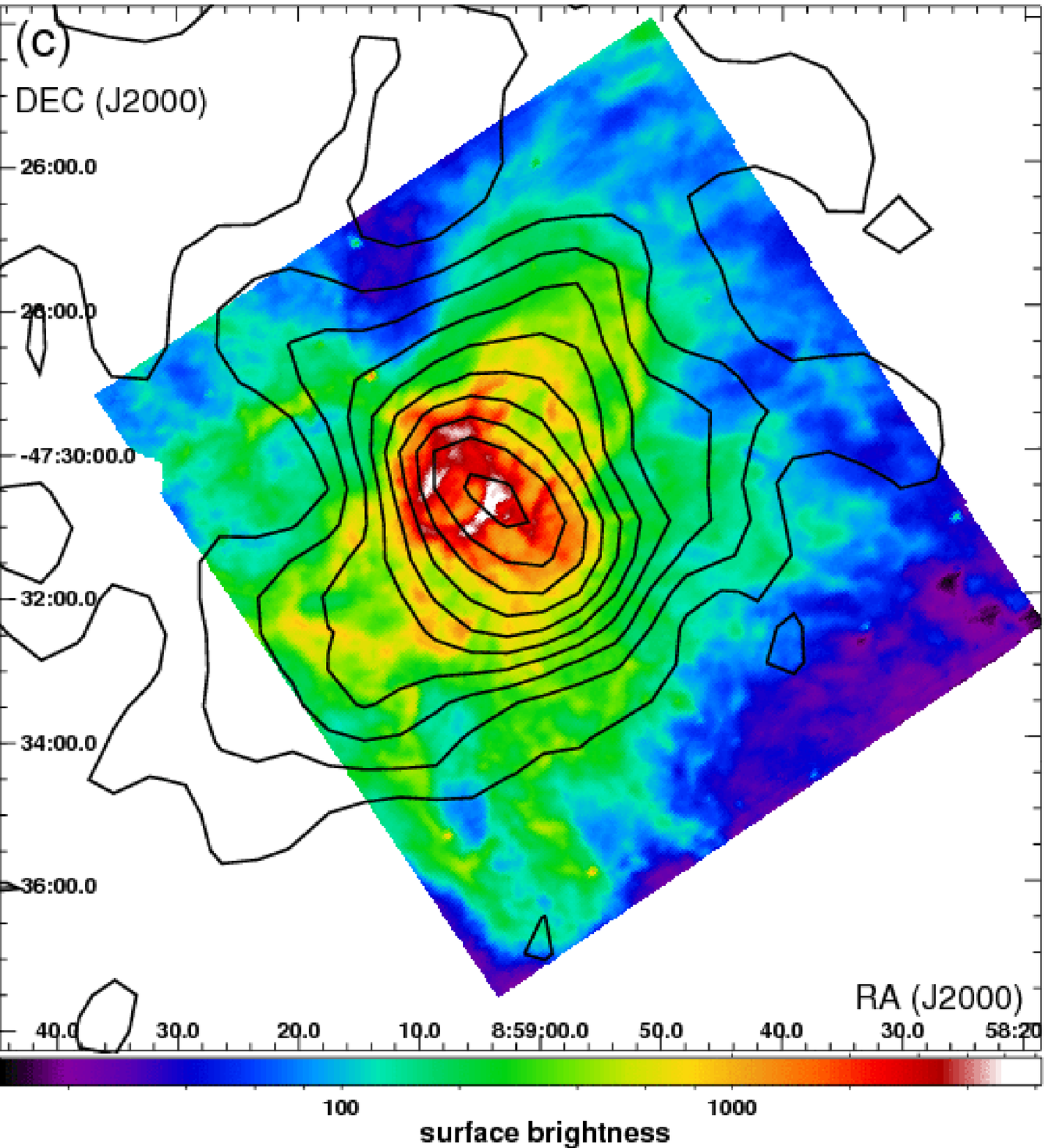}
   \includegraphics[width=6.5cm]{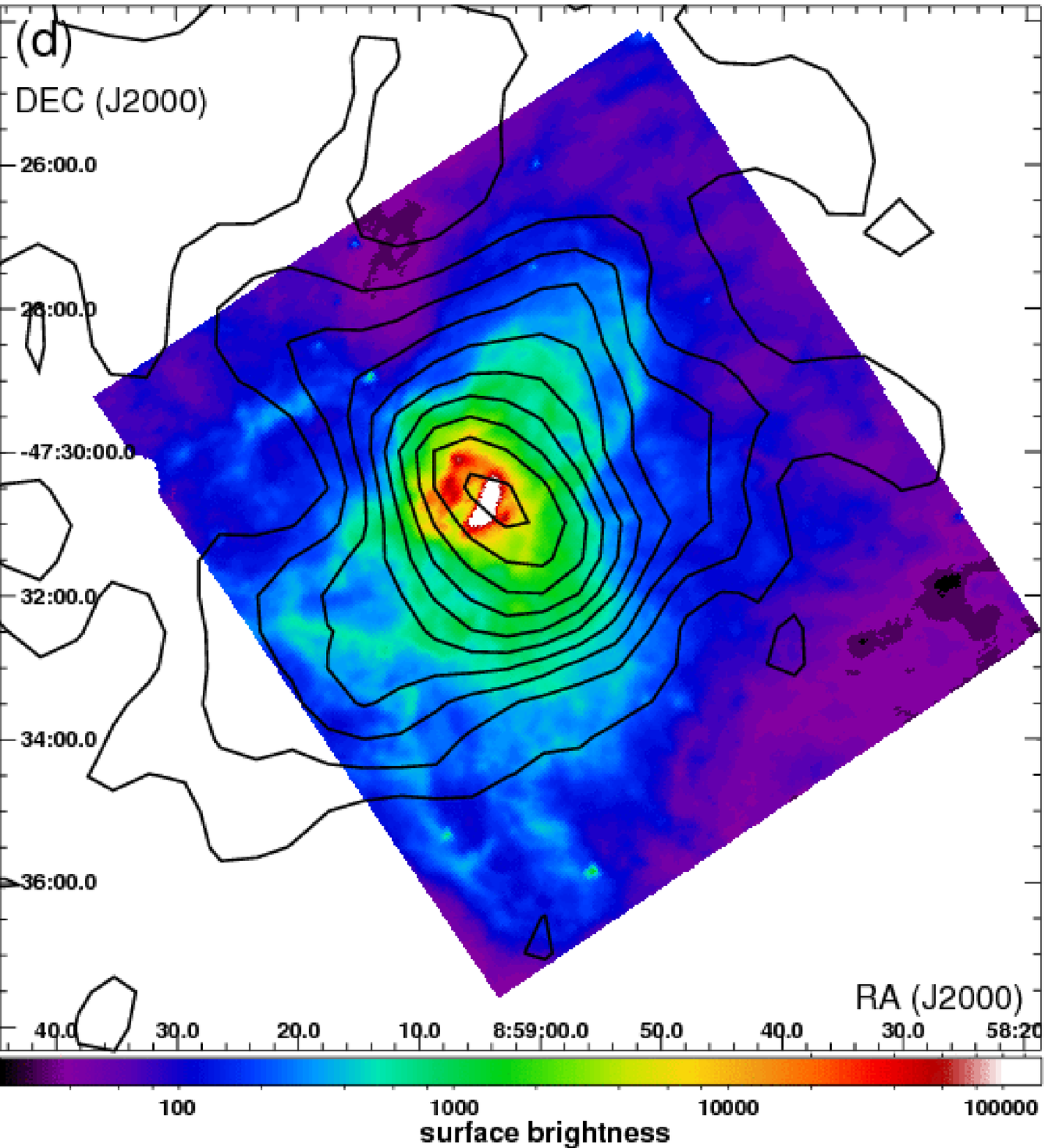}
   \caption{(a) AKARI 3 $\mu$m, (b) 4 $\mu$m, (c) 7 $\mu$m, (d) 11 $\mu$m band maps of RCW38, shown together with the [C{\small II}] contour map in Fig.1. The color levels are given in units of MJy sr$^{-1}$. The AKARI maps were created by using pointed observations.}
   \label{}
\end{figure*}

For smaller scale structures, Fig. 4 shows the images of RCW~38 obtained by the AKARI pointed observations in the {\it N3}, {\it N4}, {\it S7}, and {\it S11} narrow bands with a short exposure (\citealt{Ona07}), covering an area of $10'\times10'$ near the center. By comparing them with the [C{\small II}] contour map, there is an overall spatial correspondence between the [C{\small II}] emission and the extended emission at these bands, although their spatial resolutions are much different. Their inner structures similarly exhibit the NE--SW elongation, while their outer distributions extend toward similar directions, i.e., the north, NE, southeast (SE), and west from the center. The {\it S7} band image appears to have the closest correspondence with the [C{\small II}] distribution. In the {\it S11} band image, the intensity concentrates to a large extent relatively in the center, which is probably caused by the contribution of warm dust emission to this band, as shown later.  

To understand the extended emission in the {\it N3} and {\it N4} bands, we show in Fig.5, the 2.5--5 $\mu$m spectra of the central part of RCW~38 obtained with the AKARI/IRC. The figure shows prominent hydrogen recombination lines in the Brackett, Pfund, and Humphreys series. In addition, it is notable that the spectra show strong PAH emission at 3.3 $\mu$m and aliphatic hydrocarbon features at 3.4--3.6 $\mu$m. Hence, although the hydrogen recombination lines are included in both {\it N3} and {\it N4} bands, they contribute more to the {\it N4} band, while the PAH emission contributes only to the {\it N3} band intensity. Aside from the continuum emission, which is probably attributed to stochastically heated very small grains (\citealt{Sel84}) with some contribution of the free-free emission ($\propto\lambda^{0.1}$), the contribution of the PAH emission in the {\it N3} band is about five times larger than that of the hydrogen recombination, based on the spectra. Thus, the extended emission in the {\it N3} image is more indicative of the neutral ISM. In contrast, the emission in the {\it N4} image is representative of the ionized ISM. For comparison, we show the spectra of RCW~49, another famous massive star cluster with the distance of $\sim 5$ kpc (\citealt{Fur09}), which was taken by the IRC spectroscopic observation (table 1). The relative strengths of the hydrogen recombination lines are similar between RCW~38 and RCW~49. In contrast, the hydrocarbon features relative to the recombination lines are notably strong for RCW~38, while they are faint for RCW~49. This result demonstrates that RCW~38 is indeed an embedded cluster and indicates that a large amount of gas exists along the line of sight toward the center.     

\begin{figure*}
   \centering
   \includegraphics[width=6cm]{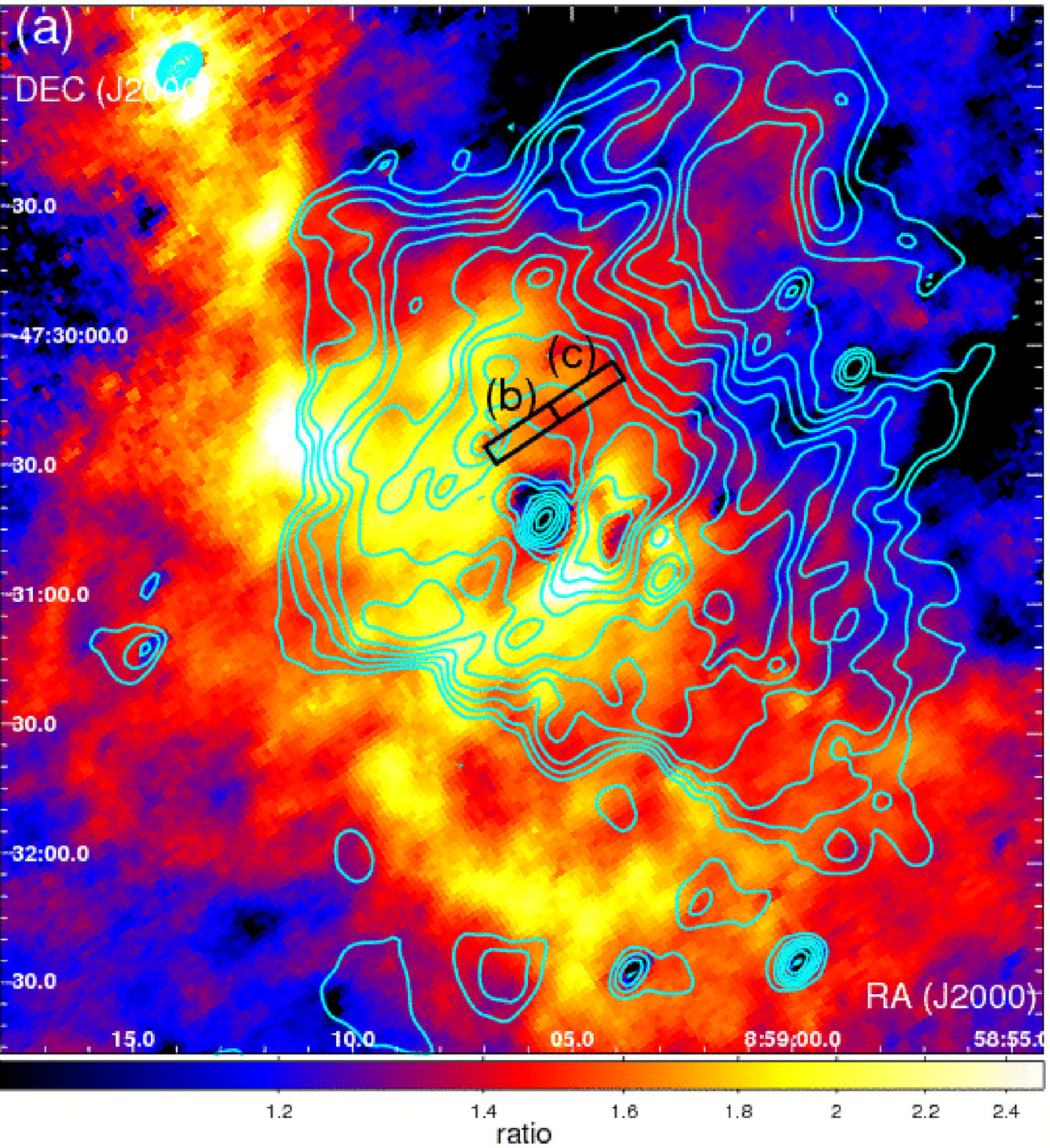}\\
   \includegraphics[width=5cm]{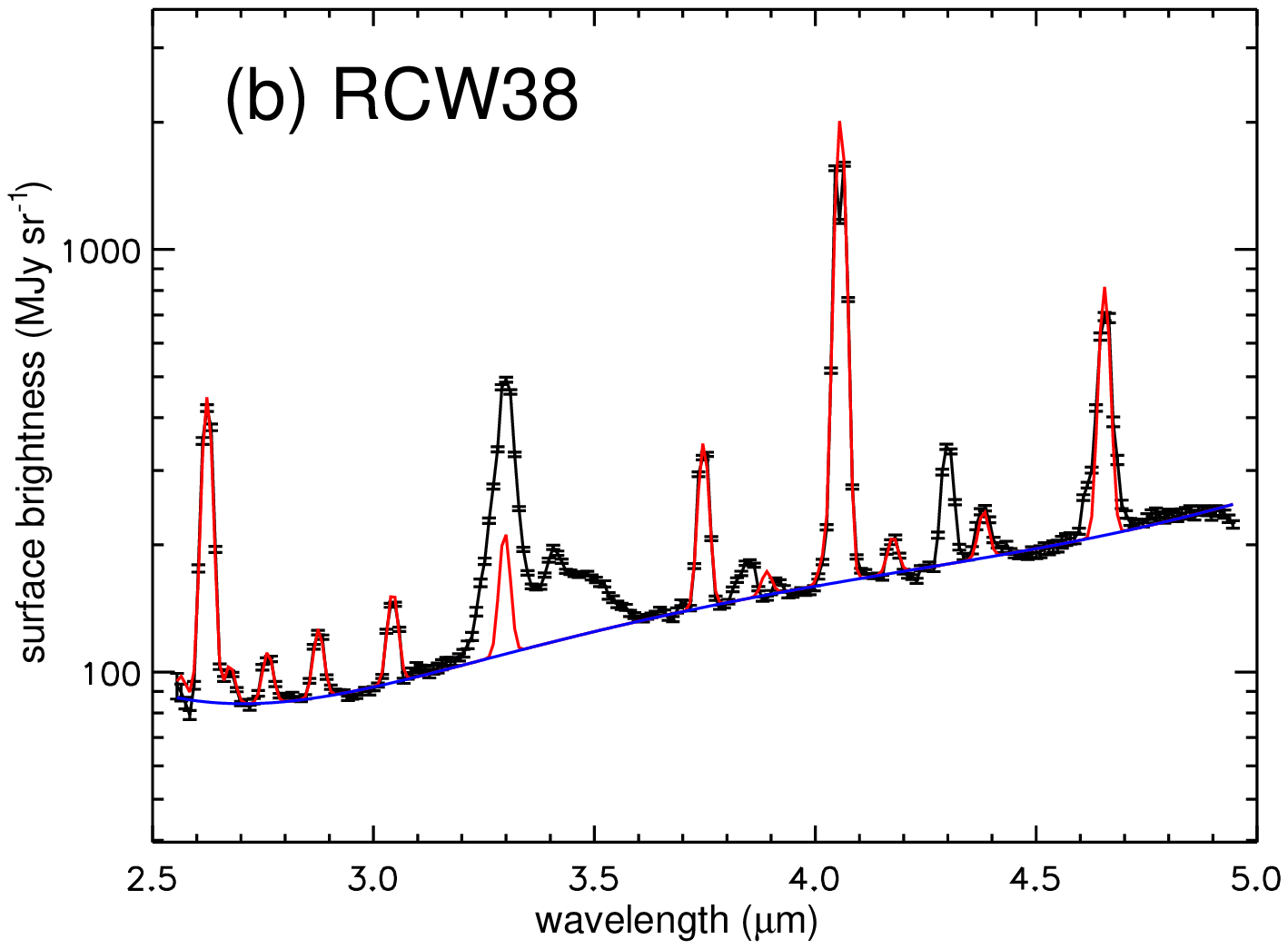}
   \includegraphics[width=5cm]{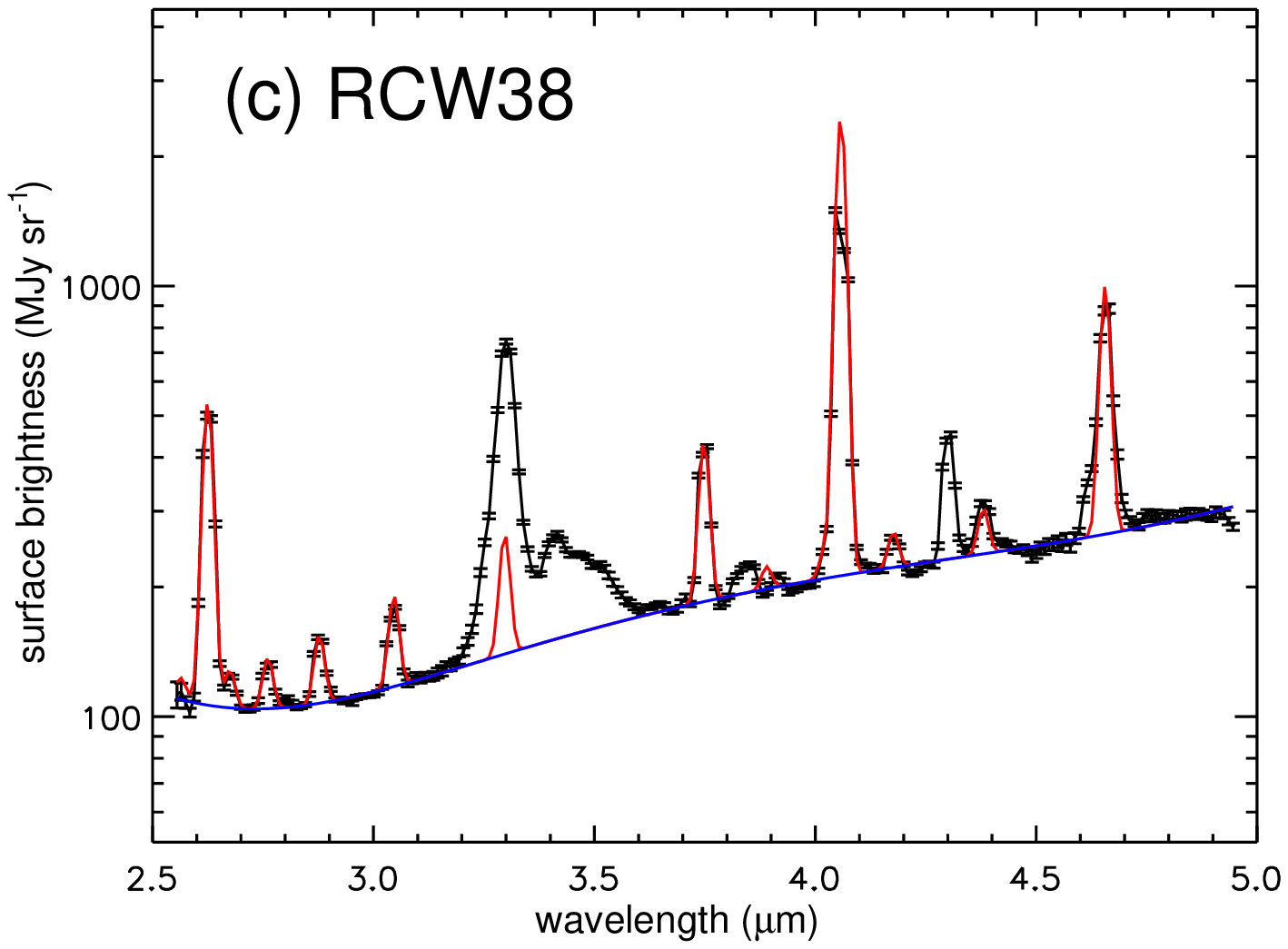}\\
   \includegraphics[width=5cm]{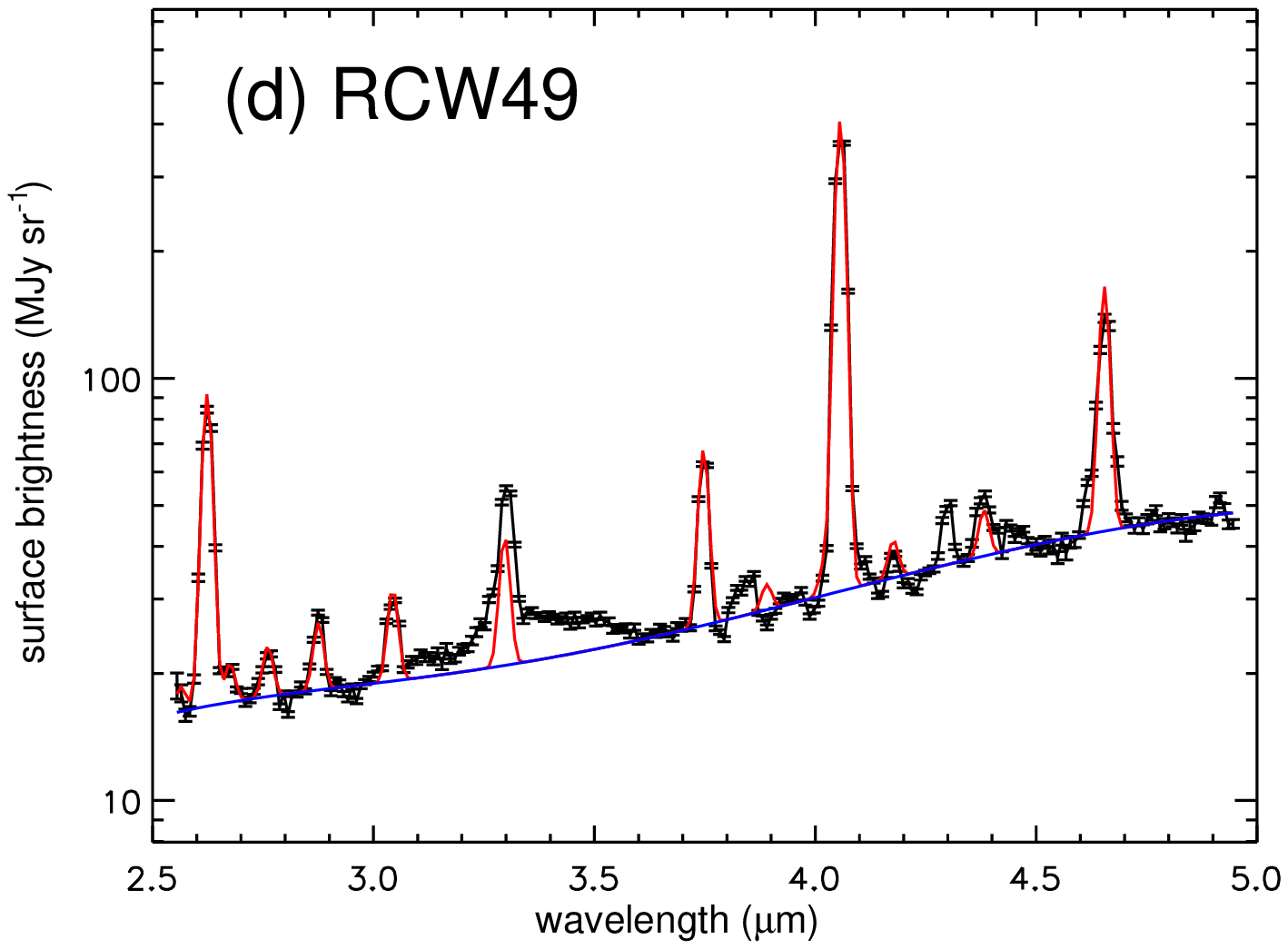}
   \includegraphics[width=5cm]{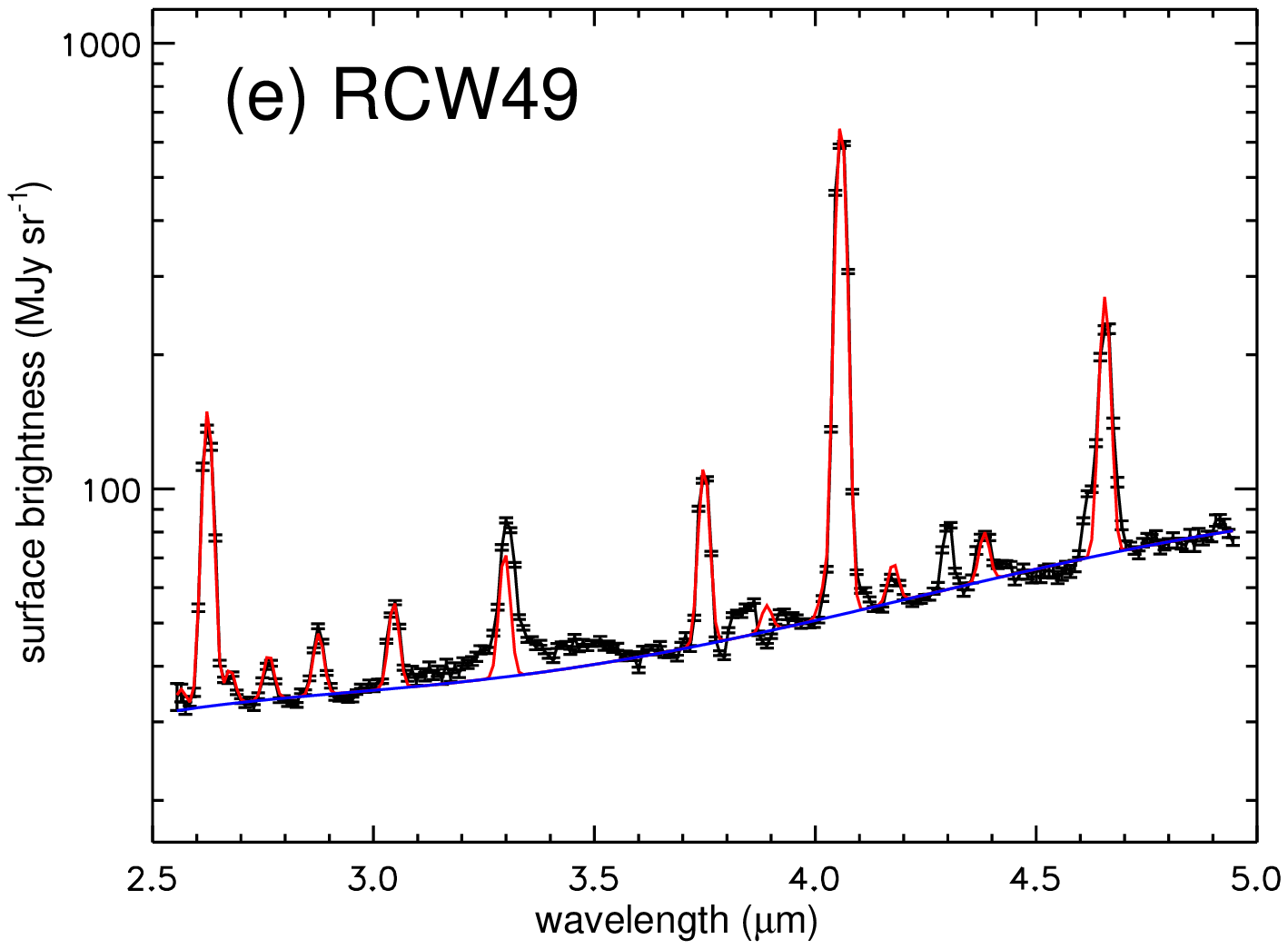}
   \caption{(a) Contour map in the AKARI 3 $\mu$m band with logarithmically spaced 13 levels from 30 to 700 MJy sr$^{-1}$, overlaid on the 4 $\mu$m to 3 $\mu$m ratio map, the same as in Fig.7a but a close-up image of the central region. The rectangles indicate the sub-slit apertures used to create the spectra in panels b and c. The AKARI 2.5--5 $\mu$m spectra of the central regions of RCW~38 (b, c), compared with those of the central regions of RCW~49 (d, e). We note that the Br$\alpha$ line at 4.05 $\mu$m is saturated in panels b and c. The red solid curve in each spectrum shows the hydrogen recombination lines estimated by the case B model calculation plus the continuum approximated by multiple blackbody components (see text for details). }
   \label{}
\end{figure*}

\section{Discussion}
To interpret the above results, we constructed near- to far-IR spectral energy distributions (SEDs) using flux densities in the {\it S9W}, {\it L18W}, {\it N60}, and {\it N160} bands from the all-sky survey and those in the {\it N3}, {\it N4}, {\it S7}, and {\it S11} bands from the pointed observations. We first removed point sources from the images using SExtractor (\citealt{Ber96}) and then smoothed the images by a common boxcar kernel of $60''$. We fitted an SED for every grid of $60''\times60''$ by a two-temperature dust modified blackbody plus PAH component model (Fig.6), excluding the {\it N3} and {\it N4} data points. We note that these components are just representatives to reproduce the shape of the SEDs and we consider below only the integrated IR intensity of each component. We need at least two dust temperatures to explain the difference in the distribution between the mid-IR and far-IR dust emissions. The PAH parameters were taken from Draine \& Li (2007) by adopting the PAH size distribution and fractional ionization typical of the diffuse ISM.  
For the modified blackbody model, we adopted an emissivity power-law index of $\beta=1$ for every component; the choice of the dust emissivity is not critical since we discuss the integrated intensities. The temperatures of cool and warm dust with initial conditions of 30 K and 150 K, respectively, and the normalizations of the three components were allowed to vary in the SED fitting. The observed [C{\small II}] line emission makes a negligible (0.3--2 \%) contribution to the {\it N160} band intensity (Fig.3b). Although the [O{\small I}] 63 $\mu$m line emission can potentially make a larger ($\sim 10$ \%) contribution to the {\it N60} band intensity (\citealt{Sal12}), it is still within the absolute flux uncertainty $\sim$20 \% of the {\it N60} band (\citealt{Kaw07}). Examples of the SED fitting are shown in Fig.6a, from which we can confirm that the PAH emission dominates in the {\it S7}, {\it S9W}, and {\it S11} bands. In very bright regions near RCW~38, however, the PAH and warm dust components contribute comparably to the {\it S11} band intensity. 
The spatial distributions of the total (2.5--1000 $\mu$m) intensities thus obtained for the PAH and the warm and cool dust components are shown in Figs.6b--6d, respectively.  

The distributions of the above three components are apparently different from each other; compared to the PAH component, the warm dust component extends more to the SE, while the cool dust component extends more to the SW. 
These differences are clearly recognized in the maps of the ratios of the warm dust and the cool dust to the PAH component in Figs.6e and 6f, respectively. In contrast, the [C{\small II}] distribution does not exhibit such biased extensions toward either SE or SW with respect to the center of RCW~38, showing a better agreement with the PAH distribution. Also, on a large scale we have already shown in Fig.2 that the [C{\small II}] emission exhibits a better spatial correspondence with the PAH emission than with the warm and cool dust emission. Therefore we confirm that the photo-electric effects of PAHs contribute significantly to the gas heating and thus the [C{\small II}] cooling in PDRs, while those of dust grains do not.

More quantitatively, considering the {\it S9W} band width (6.7--11.6 $\mu$m or $\sim$$2\times 10^{13}$ Hz; \citealt{Ona07}), the slope of the fitted line in Fig.3a corresponds to the [C{\small II}]-to-band flux ratio of $\sim$0.02, which gives a rough estimate on the [C{\small II}]-to-PAH flux ratio. Alternatively the map in Fig.6b indicates that the flux ratio of [C{\small II}] to the modeled PAH component is $\sim$0.01. The Herschel/HIFI observations of the reflection nebula NGC~7023 revealed a tight correlation between [C{\small II}] and PAH intensities along two cutting lines of lengths $1'\sim2'$ across the nebula (\citealt{Job10}). They showed a [C{\small II}]-to-PAH flux ratio of $\sim$0.01 for the [C{\small II}] intensity range of $10^{-4}-10^{-3}$ ergs s$^{-1}$ cm$^{-2}$ sr$^{-1}$. Thus our results show a fair agreement with theirs for a similar [C{\small II}] flux range. According to the calculations by \citet{Oka13}, the flux ratio of the sum of the major cooling lines ([O{\small I}] 63 $\mu$m, [O{\small I}] 145 $\mu$m, and [C{\small II}]), to PAH emission varies from 0 to 0.15, depending on the grain-charging condition. Assuming that [O{\small I}] 63 $\mu$m is comparable to [C{\small II}] and [O{\small I}] 145 $\mu$m is negligible (\citealt{Oka13}), we obtain the above flux ratio of 0.02$\sim$0.04 for RCW~38, which implies an intermediate grain-charging condition. Since Fig.3a shows no systematic deviation from a linear relation except at high [C{\small II}] intensities, it is likely that the grain-charging condition (i.e., the ionization state of PAHs) does not dramatically change within the [C{\small II}]-detectable region. At [C{\small II}] intensities higher than $10^{-3}$ ergs s$^{-1}$ cm$^{-2}$ sr$^{-1}$, [O{\small I}] is expected to become brighter than [C{\small II}] (\citealt{Hol91}) and hence a better tracer of gas cooling. This can be recognized in the deviation from a linear trend at such high intensities as in Figs.3a and 3b.

The {\it N3} and {\it N4} band flux densities were not used in the above SED fitting. Nevertheless, the N3 data points exhibit relatively good fits to the PAH 3.3 $\mu$m emission in the model. The hydrogen recombination lines were not included in the model, and therefore a significant fraction of the difference between the measurement and the prediction at 4 $\mu$m is attributed to them. The ratio map of the flux density observed at 3 $\mu$m to that predicted by the model is shown in Fig.6g. Neutral PAHs display stronger 3.3 $\mu$m emission relative to the other features than ionized PAHs do (\citealt{Dra07}). Therefore in regions with higher ratios, where neutral PAHs are dominant, few far-UV photons are likely to be available. As can be seen in the figure, such regions are extended toward the SW direction from RCW~38, which is consistent with the distribution of the CO molecular clouds revealed with NANTEN (Fig.2d).   

We calculate the ratios of surface brightness in the {\it N4} to that in the {\it N3} band in Fig.7a and the ratios of surface brightness in the {\it S11} to that in the {\it S7} band in Fig.7b. Here we subtracted the background levels measured at darkest regions within the field of view before dividing the two images. With the aid of the near-IR spectra in Fig.5, the former ratios can be interpreted as a relative importance of hydrogen recombination lines (especially Br$\alpha$) to the PAH 3.3 $\mu$m emission, thus probably tracing highly ionized regions. A region with high ratios clearly extends to the SE from the cluster, exhibiting a reasonable agreement with the distribution of the warm dust component in Fig.6e. The central part of RCW~38 exhibits a cavity structure with larger viewing angles toward the SE direction from the center, for which \citet{Smi99} showed that the winds from IRS2 have created a 0.1 pc bubble. This inner structure can also be recognized in Fig.5a, where the central peak (in the 3 $\mu$m band) and depression (in the 4 $\mu$m to 3 $\mu$m ratio) correspond to the position of IRS2. 
Hence the extension of the ionized gas and warm dust emission to the SE is likely to reflect the spatial distribution of dense gas near the central region, i.e., the cavity structure not effectively shielding UV from the cluster toward this direction, as inferred from Fig.5a. 

As discussed in Kaneda et al. (2010), the AKARI {\it S7} and {\it S11} bands can trace the ionization state of PAHs. The C-C stretching modes at 6.2 and 7.7 $\mu$m in the {\it S7} are predominantly emitted by ionized PAHs, while the C-H out-of-plane mode at 11.3 $\mu$m in the {\it S11} arises mainly from neutral PAHs (\citealt{All89}; \citealt{Job94}; \citealt{Dra07}). Therefore the contribution of neutral PAHs is expected to be relatively small in the {\it S7} band, i.e., a larger fraction of neutral PAHs for higher {\it S11}/{\it S7} ratios, except for bright regions, where the warm dust component can contribute to the {\it S11} band (Fig.6a). Hence the variations of the {\it S11}/{\it S7} ratio in Fig.7b show a trend that neutral PAHs are dominant in the SW region. This is consistent with the above observational results, such as the sharp decrease in the [C{\small II}] emission, and the presence of a cloud in the $^{12}$CO and far-IR dust emission, where a huge decrease in the number of far-UV photons is expected. Furthermore, the {\it S11}/{\it S7} ratio map shows a notable resemblance to the map of the ratio of the observed to the predicted PAH 3.3 $\mu$m emission in Fig.6g, supporting the above picture. Conversely, as can be seen in Fig.7b, there is a trend that the {\it S11}/{\it S7} ratio is suppressed to the directions in which the [C{\small II}] emission is extended, indicating a positive correlation between the PAH-ionizing photon intensity and the [C{\small II}] line intensity. Hence, although we show a tight correlation between the [C{\small II}] and PAH emission within the [C{\small II}]-detectable region, the PAH emission is extended more than [C{\small II}]. This is because photons with an energy much lower than the ionization potential of carbon ($\sim$11.3 eV) can excite PAHs and even ionize neutral PAHs if their energy is higher than $\sim$6 eV (\citealt{Mal07}). Therefore, the PAH emission is a better tracer of the whole structure of PDRs, while [C{\small II}] is more sensitive to surfaces of PDRs, likely causing the appreciable difference in their distributions under an edge-on cloud configuration in the SW region (Figs.3a and 4c). 

We fit each spectrum in Fig.5 by the hydrogen recombination lines estimated by the case B model (\citealt{Hum87}) calculation plus the continuum approximated by a fourth-degree polynomial. The line ratios are estimated from the result by \citet{Sto95} for the gas temperature and density of $10^4$ K and $10^2$ cm$^{-3}$, and the near-IR extinction curve in \citet{Lan84} is adopted for $R_{\rm V} = 3.1$. In fitting the hydrogen recombination lines, we did not use the Br$\alpha$ at 4.05 $\mu$m, which were found to be saturated in the spectra of RCW~38, or the Pf$\delta$ at 3.30 $\mu$m blended with the PAH 3.3 $\mu$m feature. As a result, we obtain $A_{\rm V} = 5.0 \pm 0.3$ mag for spectrum (a), $5.2 \pm 0.3$ mag for (b), $4.4 \pm 0.2$ mag for (c), and $5.0 \pm 0.3$ mag for (d). The former two are from RCW~38, while the latter two are from RCW~49. Hence there is no significant difference in the dust extinction between RCW~38 and RCW~49, even though the intensities of the PAH 3.3 $\mu$m feature are much different between them. The interstellar extinction toward RCW~38 and RCW~49 should contribute to a significant fraction of those $A_{\rm V}$ values. From the CO surface brightness in the central region of RCW~38, $N_{\rm H}$ and $A_{\rm V}$ are estimated to be $\gtrsim 2\times 10^{22}$ cm$^{-2}$ and $\gtrsim$ 10 mag, respectively, with the H$_2$/CO conversion factor of $1.9\times10^{20}$ cm$^{-2}$/K km s$^{-1}$ (\citealt{Nie06}) and the canonical relation $N_{\rm H}/A_{\rm V} = 2\times10^{21}$  cm$^{-2}$mag$^{-1}$ (\citealt{Boh78}). Hence the dominant CO cloud toward the center of RCW~38 does not contribute much to obscuring the hydrogen recombination lines and is thus likely located behind the HII region. Accordingly, the parental molecular cloud of the RCW~38 cluster is not likely to be the central component but present dominantly in the SW region of RCW~38. The [C{\small II}]-emitting region, by assuming its depth of $A_{\rm V}\sim 4$ mag from the cloud surface and the gas density $\sim 10^3$ cm$^{-3}$ typical of PDRs (\citealt{Hol91}), is expected to penetrate to 2.5 pc or $5'$ into the parental cloud of RCW~38. This agrees well with the observed [C{\small II}] extent to the SW direction (Fig.2).

A probable geometry of the molecular clouds with respect to the RCW~38 cluster is that one component is in the back, its surface strongly illuminated by the UV radiation from RCW~38, while another is just on the side, causing the observed rapid drop in [C{\small II}] in the SW direction. Such a complicated geometry of the clouds suggests that the formation of the RCW~38 cluster may have been triggered by a cloud-cloud collision like that of the RCW~49 cluster (\citealt{Fur09}; \citealt{Oha10}) and potentially many other massive clusters. To examine this possibility for RCW~38, a detailed study on the CO emission is needed and will be reported in a separate paper.

\begin{figure*}
   \centering
   \includegraphics[width=12cm]{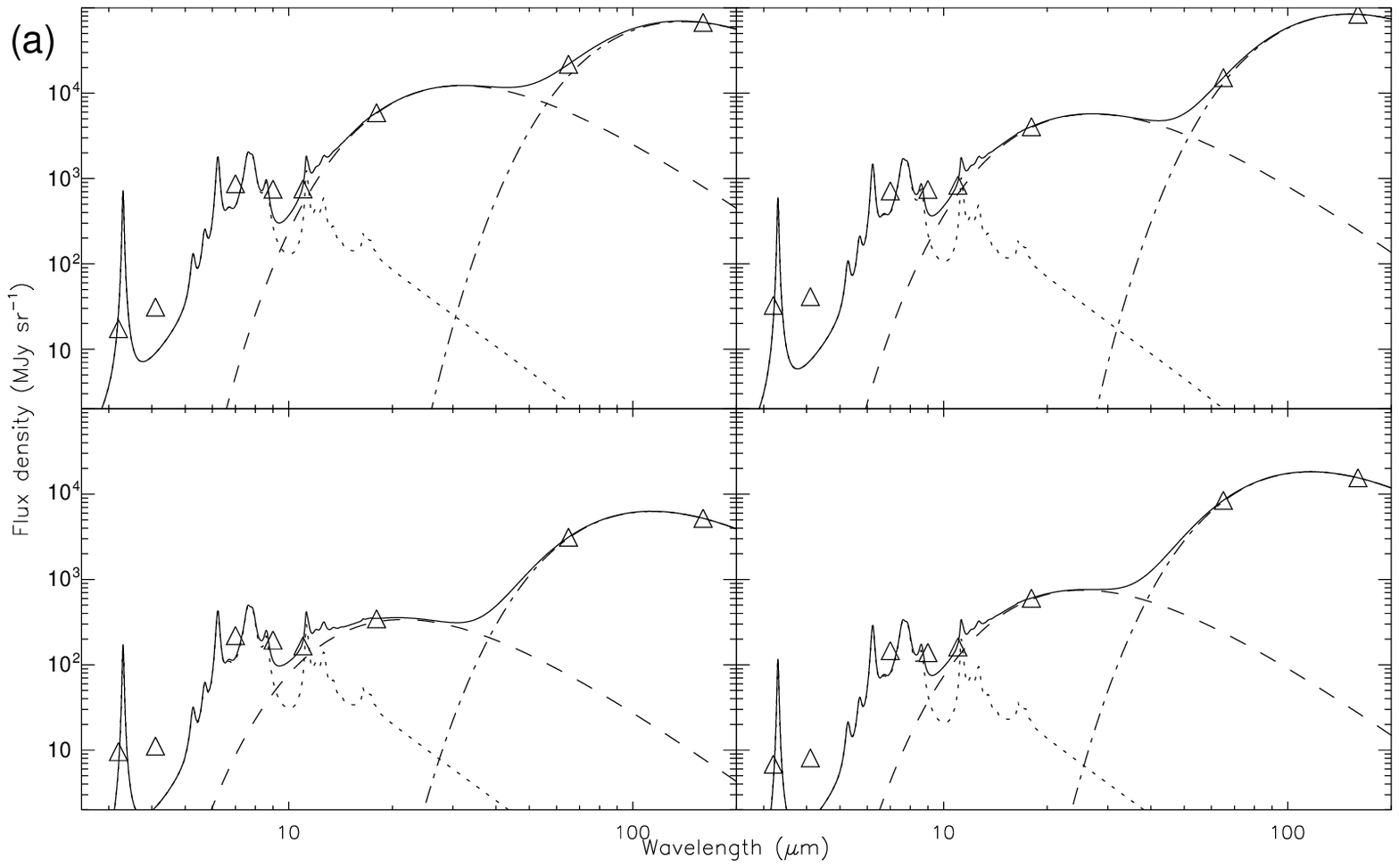}\\
   \includegraphics[width=4.3cm]{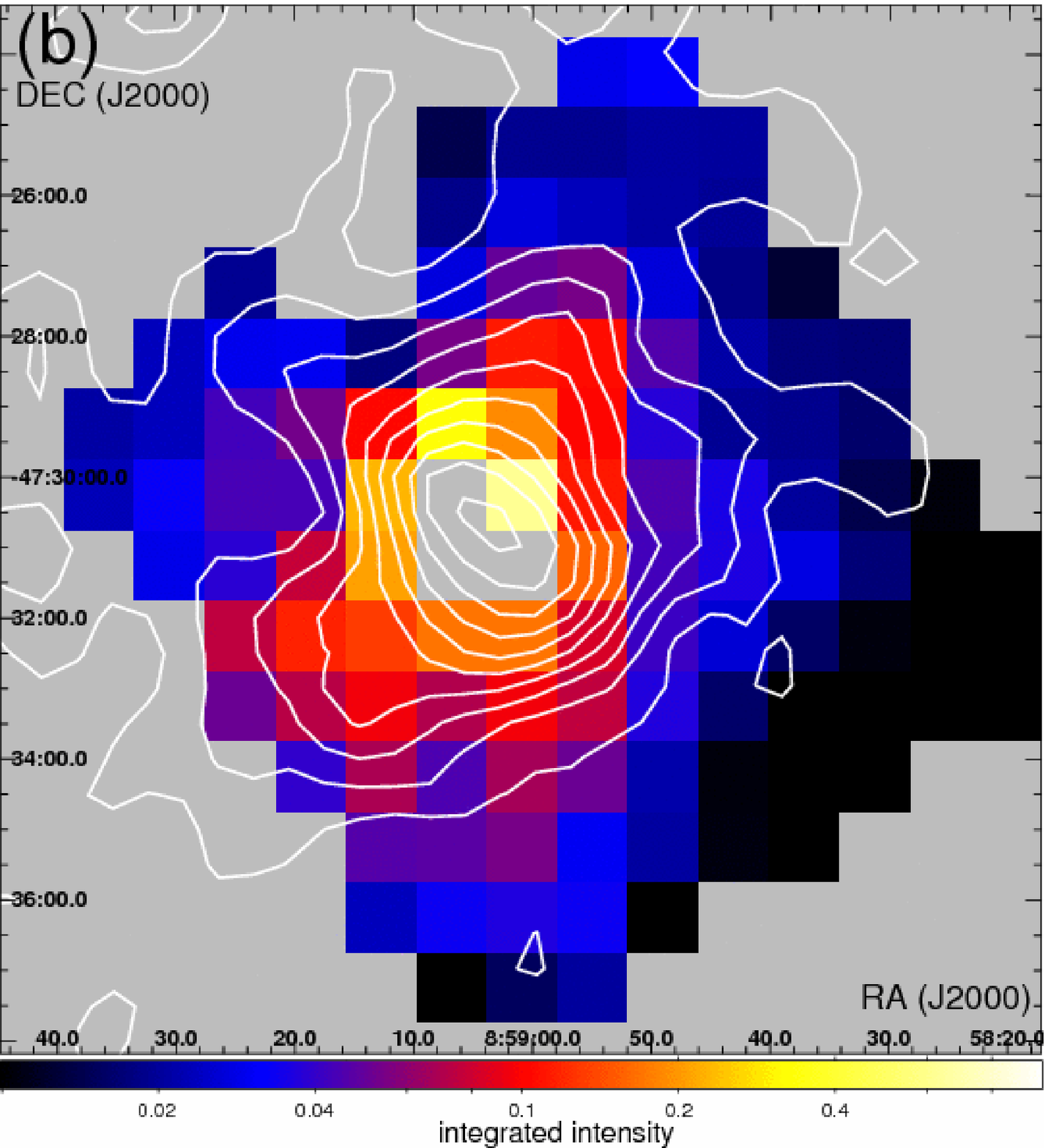}
   \includegraphics[width=4.3cm]{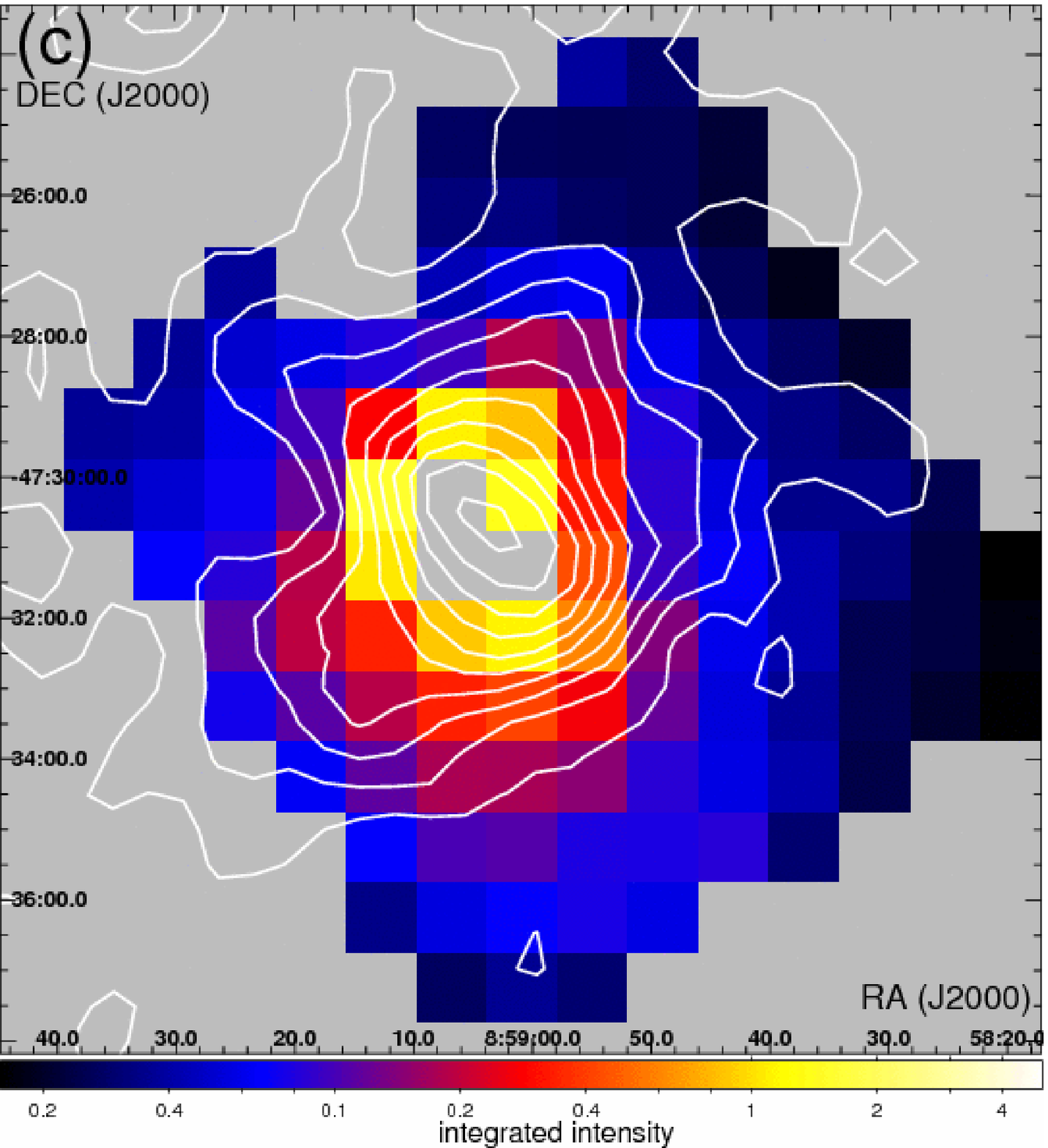}
   \includegraphics[width=4.3cm]{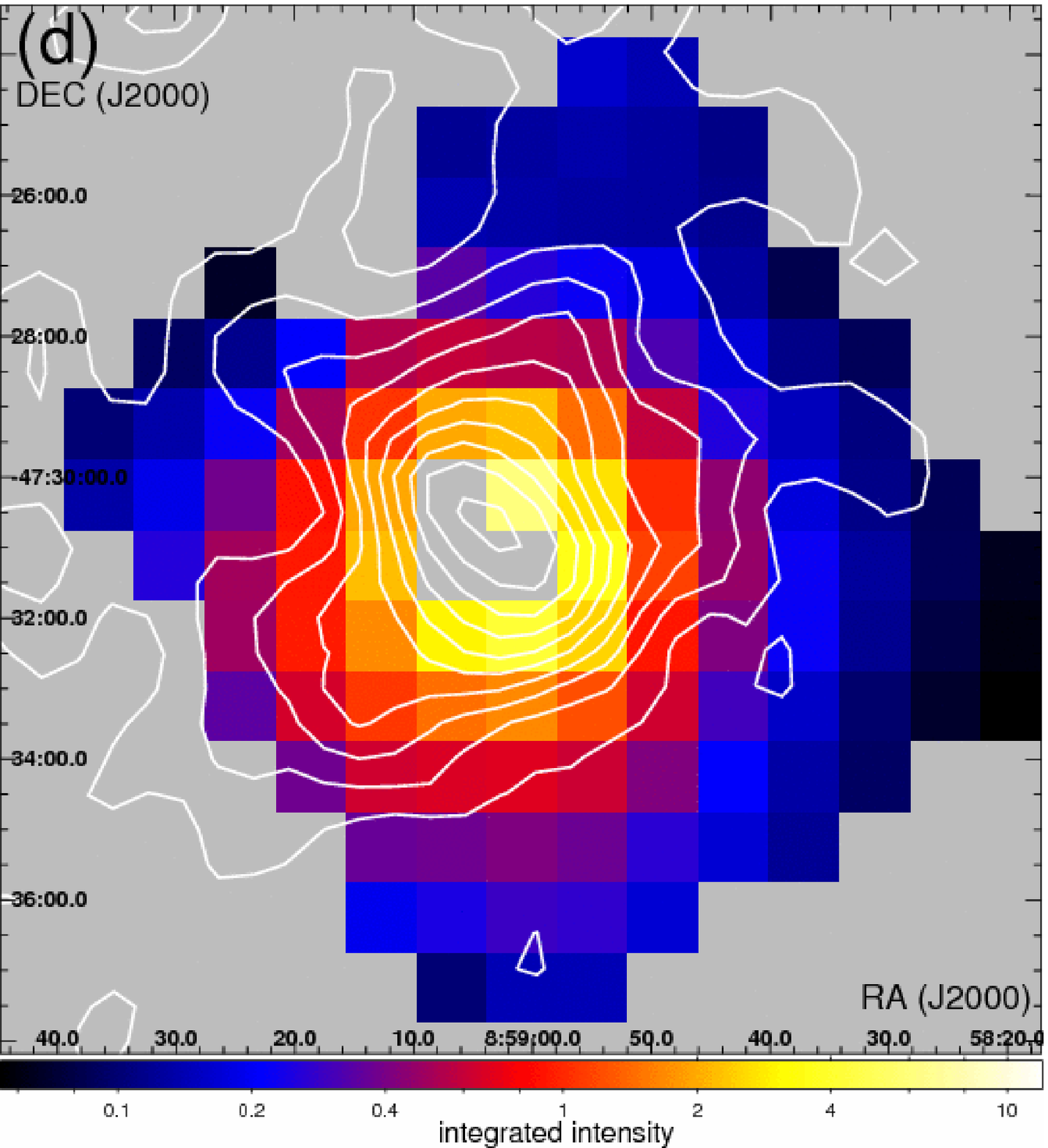}\\
   \includegraphics[width=4.3cm]{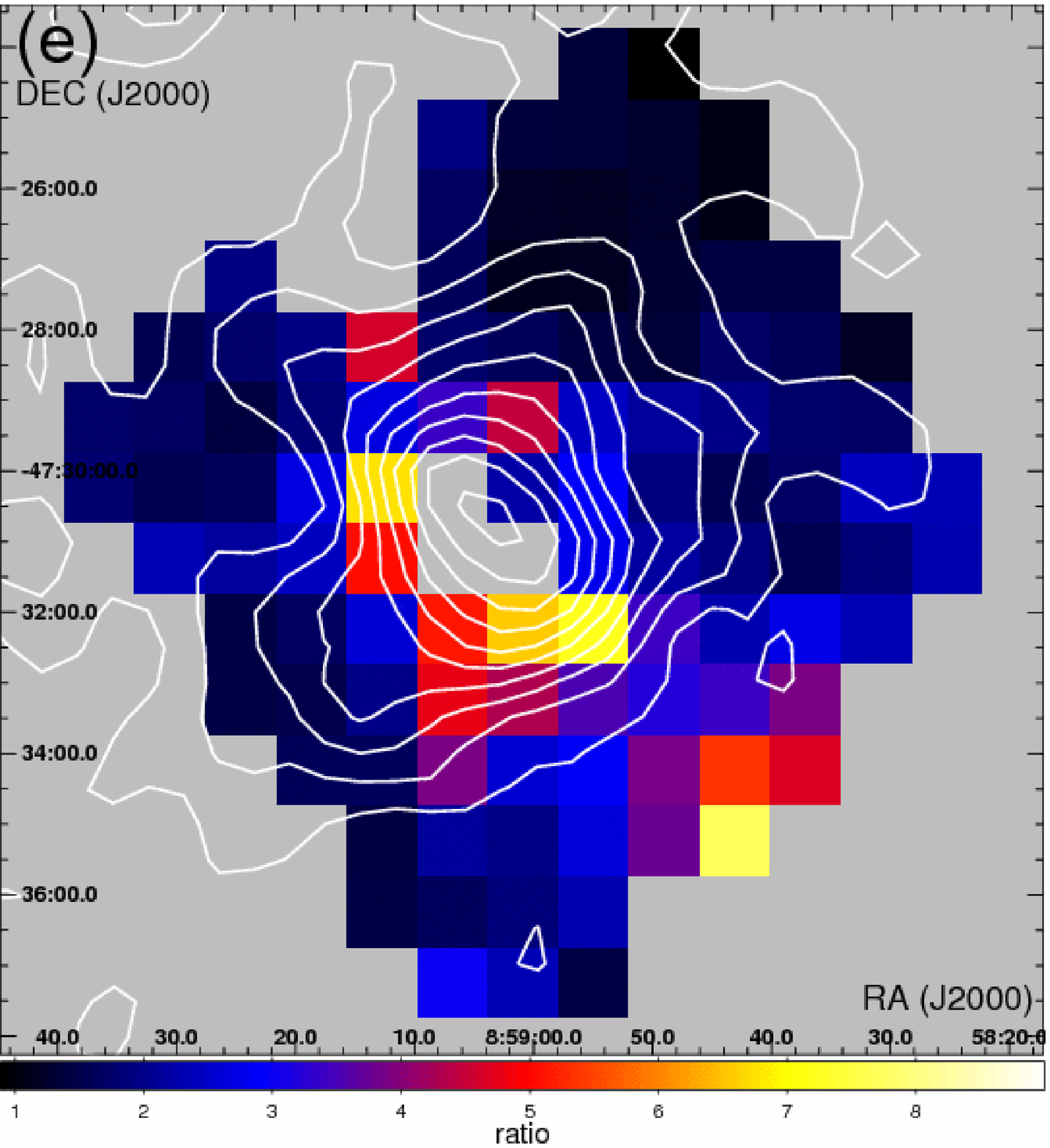}
   \includegraphics[width=4.3cm]{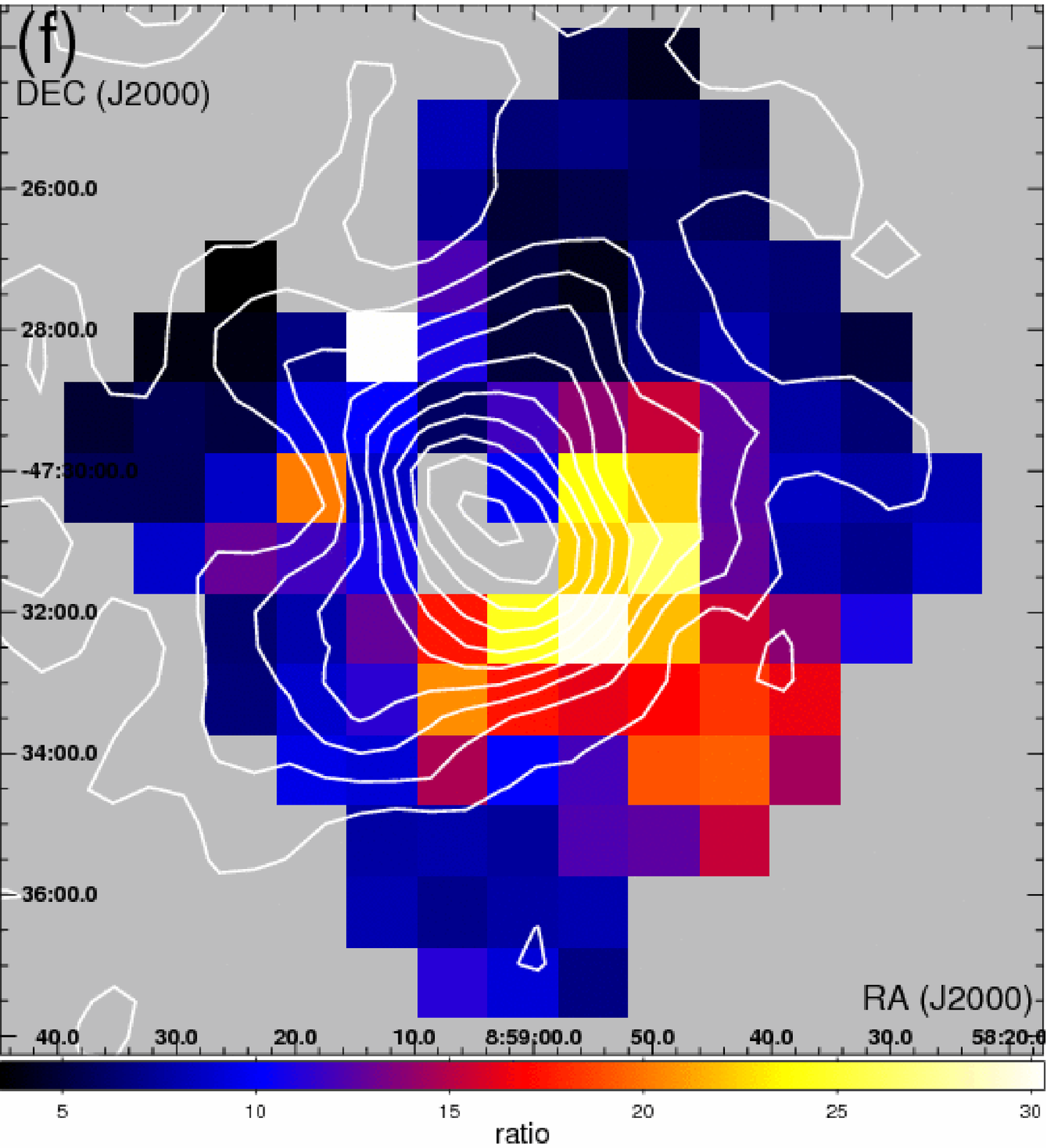}
   \includegraphics[width=4.3cm]{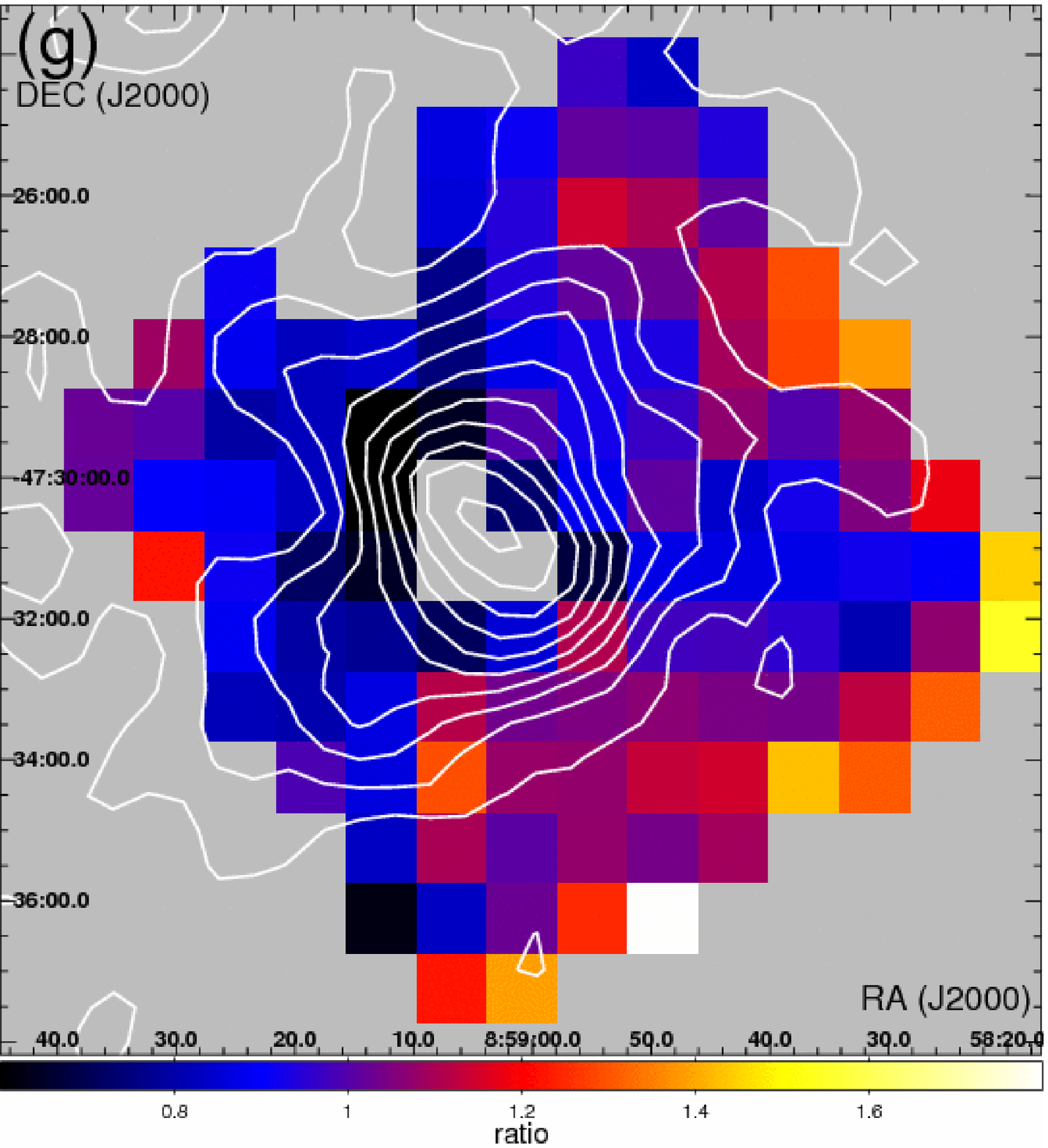}
   \caption{(a) Examples of SEDs fitted by a PAH plus double-temperature dust model. Distributions of (b) PAH, (c) warm dust, and (d) cool dust emission, decomposed by the SED fitting. The color levels correspond to the 2.5--1000 $\mu$m intensities given in units of ergs s$^{-1}$ cm$^{-2}$ sr$^{-1}$. (e) Ratios of the warm dust to the PAH and (f) the cool dust to the PAH component; the area with PAH intensity levels lower than 0.008 ergs s$^{-1}$ cm$^{-2}$ sr$^{-1}$ are masked in calculating the ratios. (g) Ratio of the observed 3 $\mu$m band intensity to the model-predicted 3 $\mu$m intensity. The superposed contours are the distribution of the [C{\small II}] emission, the same as in Fig.1.}
   \label{}
\end{figure*}

\begin{figure}
   \centering
   \includegraphics[width=6.5cm]{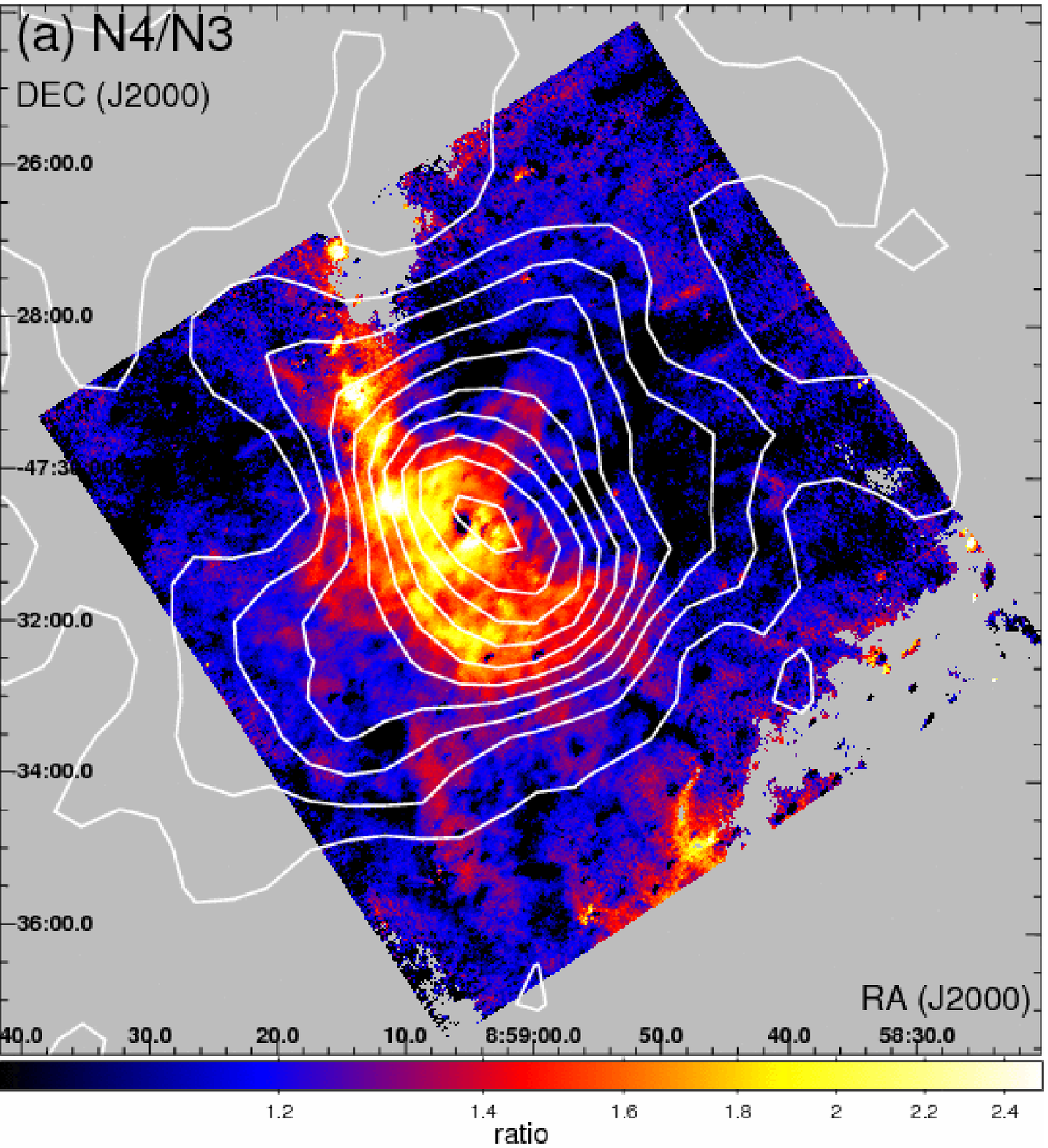}
   \includegraphics[width=6.5cm]{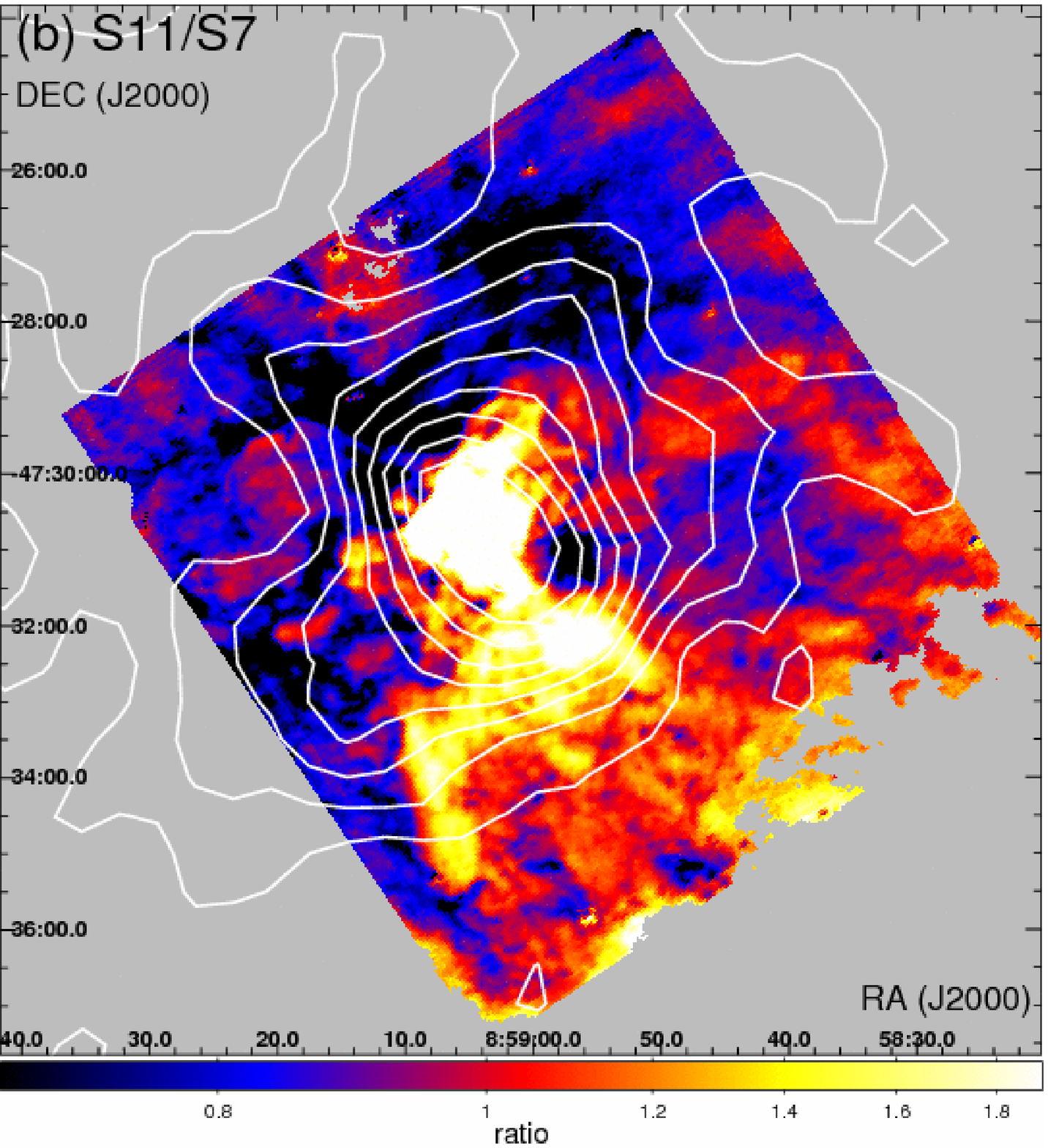}
   \caption{(a) Ratio of the 4 $\mu$m to the 3 $\mu$m band image (same as the image in Fig.5a but enlarged), and (b) that of the 11 $\mu$m to the 7 $\mu$m band image, both shown together with the [C{\small II}] contour map in Fig.1. The areas in the {\it N3} and {\it S7} band images with brightness levels lower than 1.9 MJy sr$^{-1}$ and 34 MJy sr$^{-1}$, respectively, are masked in calculating the ratios.}
   \label{}
\end{figure}

%\begin{figure}
%   \centering
%   \includegraphics[width=8cm]{../image/nanten-rcw38.ps}
%   \caption{[C{\small II}] line intensity map in the white contours, overlaid on the composite color map of the AKARI 9 $\mu$m and 18 $\mu$m band intensities in blue and red, respectively. NANTEN CO J=1-0 contour maps of the velocity ranges of $-5\sim+2$ km s$^-1$ in the cyan contours and $+7\sim+15$ km s$^-1$ in the pink contours are shown together, where the contour levels are commonly drawn from 7 to 40 K km$^{-1}$ sin 7 steps. }
%   \label{}
%\end{figure}

\section{Summary}
We performed large-scale mapping of RCW~38 in the [C{\small II}] 158 $\mu$m line with the balloon-borne telescope and in the PAH and dust emission with AKARI to understand their interplay as gas cooling and heating agents as well as the geometry of the associated molecular clouds. We find that the [C{\small II}] emission is widely extended around the cluster, exhibiting the structure extended to the north and east from the center with a rapid decline toward the SW direction. On a smaller spatial scale, the [C{\small II}] emission is extended to the north, NE, SE, and west from the center.  The distribution of the [C{\small II}] emission follows that of the PAH emission very well on both large and small spatial scales, better than that of the dust emission, confirming the relative importance of PAHs for photo-electric heating of gas in PDRs. In the SW region, neutral PAHs, cool dust, and CO molecular cloud are dominant, all consistently suggesting a huge decrease in the number of far-UV photons there. We find that the molecular cloud toward RCW~38 only fractionally obscures the hydrogen recombination lines from RCW~38, indicating that the central cloud with the peaks of the [C{\small II}] and PAH emission is located behind RCW~38. The star cluster is not embedded in this cloud component, and another component, likely a parental cloud, is dominantly lying in the SW region, shielding the UV from RCW~38 to suppress the [C{\small II}] emission in the cloud.

\begin{acknowledgements}
We thank the anonymous referee for giving us useful comments and suggestions. We express many thanks to all the members of the Infrared Astronomy Group of TIFR and the members of the TIFR Balloon Facility in Hyderabad, India, for their support during the balloon campaign. Part of this work is based on observations with AKARI, a JAXA project with the participation of ESA. This research is supported by a Grant-in-Aid for Scientific Research No. 22340043 from the Japan Society for the Promotion of Science (JSPS), the India-Japan cooperative science program from JSPS and the Department of Science and Technology, the Government of India, and the Nagoya University Global COE Program, ``Quest for Fundamental Principles in the Universe: from Particles to the Solar System and the Cosmos'', from the Ministry of Education, Culture, Sports, Science and Technology of Japan. 
\end{acknowledgements}

\end{document}